\documentclass[aps,prd,reprint,superscriptaddress,nofootinbib,floatfix,longbibliography]{revtex4-1}
\usepackage[dvipsnames]{xcolor}
\usepackage{amssymb,amsmath,amsfonts,bm,slashed,soul,ragged2e,graphicx,epstopdf,hyperref,array,float}
\usepackage[utf8]{inputenc}
\usepackage{colortbl,color,caption,subcaption}
\usepackage{makecell}
\enlargethispage{\baselineskip}
\definecolor{steelblue}{RGB}{25,25,112}
\definecolor{dullblue}{rgb}{0,0.298,0.49}
\definecolor{darkred}{rgb}{0.545,0,0}
\definecolor{darkorange}{RGB}{222,132,69}
\definecolor{darkgreen}{RGB}{126,171,85}
\definecolor{blue2}{cmyk}{1, 0.1, 0.1, 0}
\hypersetup{colorlinks,linkcolor={darkred},citecolor={blue},urlcolor={dullblue}}
\DeclareCaptionJustification{justified}{\justifying}
\everymath{\displaystyle} 
\allowdisplaybreaks

\newcommand{\be}{\begin{equation}}
	\newcommand{\ee}{\end{equation}}
\newcommand{\ba}{\begin{eqnarray}}
	\newcommand{\ea}{\end{eqnarray}}
\newcommand{\mi}{\mathrm{i}}
\newcommand{\D}{\mathrm{d}}

\definecolor{seagreen}{rgb}{0.180392,0.545098,0.341176}

\begin{document}

\title{Eliminating Incoherent Noise: A Coherent Quantum Approach in Multi-Sensor Dark Matter Detection}

\author{Jing Shu}
\affiliation{School of Physics and State Key Laboratory of Nuclear Physics and Technology, 
	Peking University, Beijing 100871, China}
\affiliation{Center for High Energy Physics, Peking University, Beijing 100871, China}
\affiliation{Beijing Laser Acceleration Innovation Center, Huairou, Beijing, 101400, China}
\author{Bin Xu}
\affiliation{School of Physics and State Key Laboratory of Nuclear Physics and Technology, 
	Peking University, Beijing 100871, China}
\author{Yuan Xu}
\affiliation{International Quantum Academy, Shenzhen 518048, China}
\affiliation{Shenzhen Branch, Hefei National Laboratory, Shenzhen 518048, China}

\begin{abstract}

We propose a novel dark matter detection scheme by leveraging quantum coherence across a network of multiple quantum sensors. This method effectively eliminates incoherent background noise, thereby significantly enhancing detection sensitivity. This is achieved by performing a series of basis transformation operations, allowing the coherent signal to be expressed as a combination of sensor population measurements without introducing background noise. We present a comprehensive analytical analysis and complement it with practical numerical simulations. These demonstrations reveal that signal strength is enhanced by the square of the number of sensors, while noise, primarily due to operational infidelity rather than background fluctuations, increases only linearly with the number of sensors. Our approach paves the way for next-generation dark matter searches that optimally utilize an advanced network of sensors and quantum technologies.

\end{abstract}

\maketitle
\section{Introduction}
\label{sec:intro}

The endeavor to comprehend dark matter (DM)~\cite{bertone_2010,Salucci_2019} is pivotal in modern astrophysics and fundamental physics. Although various observations validate its existence~\cite{doi:10.1146/annurev.astro.39.1.137, Massey_2010, Markevitch_2004}, its fundamental nature remains elusive. Recently, attention has moved towards sub-eV mass bosons~\cite{Arias:2012az, jackson2023search}, including quantum chromodynamics axions~\cite{Preskill:1982cy, Abbott:1982af, Dine:1982ah}, axion-like particles~\cite{Graham:2015cka, Co:2020xlh}, and dark photons~\cite{holdom1986two, Nelson:2011sf, Graham:2015rva} as potential DM candidates. These candidates, predicted by theories involving higher-dimensional space compactifications~\cite{Cvetic:1995rj, Svrcek:2006yi, Abel:2008ai, Arvanitaki:2009fg, Goodsell:2009xc}, exhibit a ``wave nature"~\cite{Lin:2018whl, Centers:2019dyn} due to their large de Broglie wavelength and high local occupation number, allowing them to be treated as coherently oscillating fields within their correlation time and distance.

Electromagnetic interactions are crucial for detecting DM fields~\cite{Holdom:1986ag,Sikivie:1983ip}. These interactions can generate detectable currents within sensors under strong magnetic fields for axions, or directly influence electromagnetic fields for dark photons. Techniques such as resonant microwave cavities~\cite{Sikivie:1983ip, Sikivie:1985yu, Wagner:2010mi, haystaccollaboration2023new} and lumped-element circuits~\cite{Sikivie:2013laa, Chaudhuri:2014dla, Kahn:2016aff, Salemi:2021gck} have been utilized to detect these subtle signals. However, these methods contend with significant challenges from overwhelming thermal and quantum noise, which restricts detection sensitivity by introducing environmental background noise and the Standard Quantum Limit (SQL)~\cite{braginsky1967classical} noise. Recently, innovative experiments have begun employing novel quantum detection technologies to surpass the SQL in DM detection. These innovations include the use of squeezed states~\cite{HAYSTAC:2020kwv}, mode entanglement and state swapping techniques~\cite{Chen:2021bgy, Wurtz:2021cnm, Chen:2023ryb, Jiang:2022vpm}, and single microwave photon detectors~\cite{Dixit:2020ymh, Agrawal:2023umy}, facilitated by quantum non-demolition (QND) measurements~\cite{RevModPhys.52.341, RevModPhys.68.1}.

When a network or array of quantum sensors operates within the correlation time and distance of the DM field, the detection sensitivity and scan rate can be significantly improved ~\cite{Chen:2021bgy, Brady:2022bus, Brady:2022qne, Chen:2023swh, Ito:2023zhp}, incorporating a proper set of quantum operations. In particular, an enhancement of the signal rate by a factor of $O(N^2)$ can be achieved with complex entangled states of $N$ quantum sensors, while the noise scales linearly with $N$. As each detector coherently excites the DM signals, these signals can be interpreted as a combination of sensor population measurements, though they remain constrained by incoherent background noise. Importantly, the quantum coherence between different sensors, which is entirely absent in uncorrelated noise, represents a valuable resource for DM detection and merits further exploration in this area.

\begin{figure}[]
\centering
\includegraphics[width=0.48\textwidth]{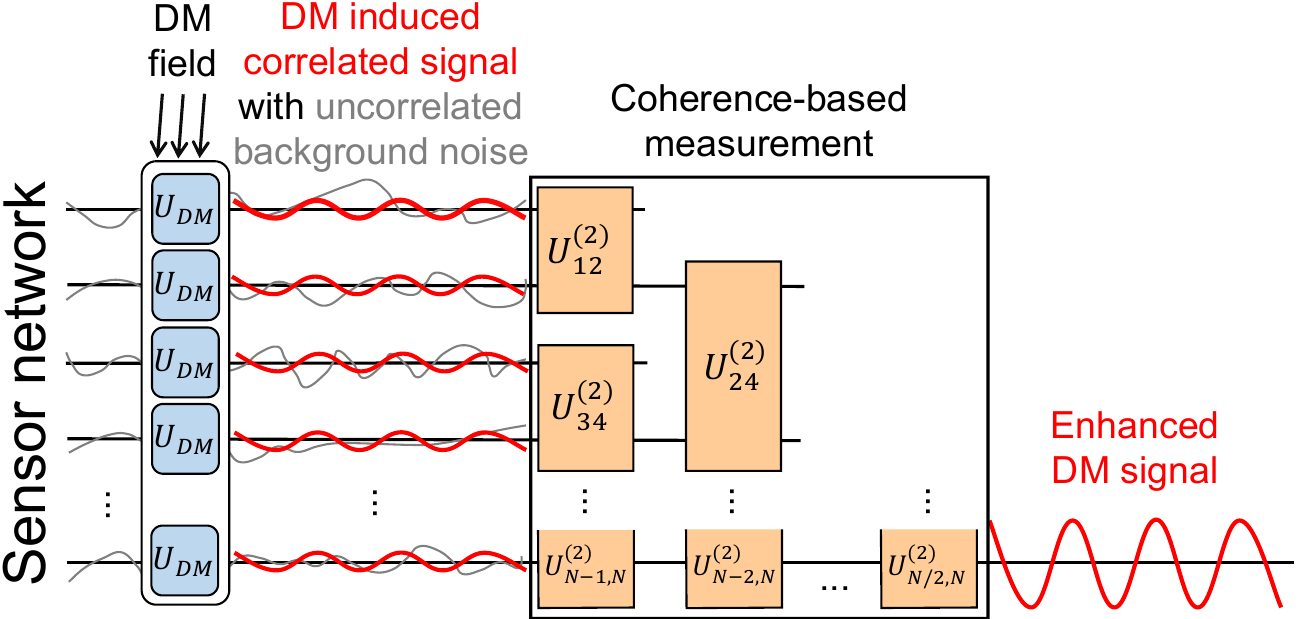}
\caption{Illustration of the protocol for coherence-based detection of the DM signal, where $U_{DM}$ represents the influence of the DM field, and $U^{(2)}_{IJ}$ denotes the transformation of the coherence signal between a pair of sensors into population numbers. Ultimately, we achieve a signal strength enhanced by the square of the number of sensors, with background fluctuations effectively canceled.
}
\label{fig:img0}
\end{figure}

In this work, we propose a novel approach as depicted in Fig.~\ref{fig:img0} that utilizes quantum coherence between DM sensors as an additional degree of freedom, a feature absent in classical systems. By integrating QND measurements to eliminate quantum errors from state preparation and measurement (SPAM), we can, in principle, achieve a noise-free coherence channel. The coherent DM signals can be extracted by measuring the photon number populations through a series of basis transformation operations. Because the noise introduced by transformation operations is considerably less than the noise from conventional measurements, we achieve an improved signal-to-noise ratio (SNR), opening up new possibilities for DM searches with enhanced detection rates and reduced noise levels.

\section{Dark matter detection with a single quantum sensor}\label{sec:single}

The interaction Hamiltonian describing the coupling between a sensor mode and the background bosonic DM field in the interaction picture is expressed as:
\begin{equation}\label{eq:Ha}
{H}_{\alpha} = {\alpha \Psi \left(\hat{a} e^{\mi \omega_r t} + \hat{a}^\dagger e^{-\mi \omega_r t}\right)}/{\sqrt{2}},
\end{equation}
where $\alpha$ denotes the coupling constant, $\hat{a}$ ($\hat{a}^\dagger$) is the annihilation (creation) operator of the sensor mode, and $\omega_r$ is the resonant frequency of mode $\hat{a}$. The oscillating bosonic DM field is represented by:
\begin{equation}
\Psi = \Psi_0 \cos(m_\Psi t + \phi_\Psi),
\end{equation}
where $\Psi_0$ is the field amplitude, $m_\Psi$ is the mass of the boson corresponding to its Compton frequency, and $\phi_\Psi$ is a random phase. Further details on the DM field model are provided in the Supplemental Material (SM).

For weak signals on resonance ($\omega_r = m_\Psi$), the quantum sensor, initialized in its ground state $|g\rangle$, evolves according to the Hamiltonian in Eq.~\eqref{eq:Ha} into the state
\begin{equation}\label{beta}
|\beta\rangle \equiv U_{DM}(\beta)|g\rangle \simeq \left(1 - \frac{|\beta|^2}{2}\right)|g\rangle + \beta|e\rangle,
\end{equation}
which is a superposition of the ground state $|g\rangle$ and the excited state $|e\rangle$. Here, $U_{DM}(\beta)=e^{-iH_{\alpha}\tau}=e^{\beta\hat a^\dagger-\beta^*\hat a}$, and $\beta = -\frac{i}{2\sqrt{2}} \alpha \Psi_0 e^{i \phi_\Psi} \tau$ represents the displacement amplitude with $|\beta|^2 \ll 1$, and $\tau < \min\{T_\Psi, T_s\}$ is the signal integration time, which is shorter than both the coherence time of the DM field $T_\Psi$ and that of the sensor $T_s$. Terms of higher order than $|\beta|^2$ in the expansion have been neglected.

The DM signal in an ideal quantum sensor is quantified by the excitation probability, defined by the expectation value of the observable $\mathcal{O}_P \equiv |e\rangle\langle e|$:
\begin{equation}
\label{eq:Pe}
    P_e = \langle \beta|\mathcal{O}_P|\beta\rangle \simeq |\beta|^2.
\end{equation}

However, in reality, detection is impeded by various types of noise. Major noise sources during the detection process include decoherence due to thermal relaxation/excitation and dephasing, along with SPAM errors. As demonstrated in Refs.~\cite{Dixit:2020ymh, Agrawal:2023umy}, SPAM errors can be significantly reduced to negligible levels by employing a repeated QND measurement and a voting protocol. We now focus on the challenges posed by thermal decoherence and dephasing errors.

In order to properly represent the ensemble average over the noise, it is recommended to utilize the quantum density matrix of the qubit
\begin{equation}
\begin{array}{ccc}
\rho=|\beta\rangle\langle\beta|\simeq & \begin{pmatrix}
        1-|\beta|^2 & \beta^*\\
        \beta & |\beta|^2
    \end{pmatrix} & \begin{matrix}
\end{matrix}
\end{array},
\end{equation}
where $\rho_{ij}=\langle i|\rho|j\rangle$ is the density matrix element.

The effect of the decoherence errors can be expressed as
\begin{equation}\label{eq:deco}
\rho\rightarrow \mathcal{E}(\rho)=\begin{pmatrix}
p_{T_1}\rho_{gg}+p_{\rm reset}p_0 & p_{T_2}\rho_{ge}\\
p_{T_2}\rho_{eg} & p_{T_1}\rho_{ee}+p_{\rm reset}p_1
\end{pmatrix},
\end{equation}
where $p_{T_1}=e^{-\tau/T_1}$ and $p_{T_2}=e^{-\tau/T_2}$ are the probability for a qubit to relax and dephase during an integration time $\tau$,
$T_1$ and $T_2$ are the thermal relaxation and dephasing times, 
$p_0$ and $p_1=1-p_0$ are the populations of the ground and excited state at equilibrium, and $p_{\rm reset}=1-p_{T_1}$ is the probability for a qubit to reset to an equilibrium state.

From Eq.~\eqref{eq:Pe}, the excitation probability in the presence of noise is given by:
\begin{equation}
\text{Tr}[\mathcal{E}(\rho)\cdot\mathcal{O}_P]=p_{T_1}|\beta|^2 + p_{\text{reset}}p_1,
\end{equation}
where the signal is suppressed by a factor of $p_{T_1}$, and a noise level of $p_{\rm reset} p_1$ is added. The SNR is given by
\begin{equation}
\label{eq:SNR1}
    SNR=\frac{S}{B}=\frac{p_{T_1}|\beta|^2}{p_{\rm reset}p_1},
\end{equation}
where $S$ and $B$ represents the signal strength and background noise, respectively. The signal significance denoted by the signal over the fluctuation of the background is given by
\begin{equation}\label{eq:sig1}
    \frac{S}{\sqrt{B}}=\frac{p_{T_1}|\beta|^2}{\sqrt{p_{\rm reset}p_1}}.
\end{equation}

\section{Coherence-based detection with multiple quantum sensors}\label{sec:series}

We intend to use multiple quantum sensors operating simultaneously to improve detection sensitivity. As long as the sensors are situated within the DM coherence length, they will experience coherent excitations. This contrasts with noisy excitations, which are inherently incoherent. For simplicity, let us consider two identical sensors, $A$ and $B$, that operate independently without interactions between them and carry the same signal strength. However, the noise affecting each sensor may differ. 
 
The density matrix of the entire system can be expressed as a tensor product:
\begin{align}
\rho^{(2)} = \mathcal{E}_A(\rho^{(1)}) \otimes \mathcal{E}_B(\rho^{(1)}),
\end{align}
where $\rho^{(1)} = |\beta\rangle\langle\beta|$ is the density matrix of a single sensor, either $A$ or $B$, and decoherence errors affect each independently. The observable for measuring the total excitation in each sensor is represented by the operator $\mathcal{O}^{(2)}_P \equiv |e_Ag_B\rangle\langle e_Ag_B| + |g_Ae_B\rangle\langle g_Ae_B|$, where the subscript $P$ indicates that the operator measures population of the excited states. The expectation value of $\mathcal{O}^{(2)}_P$ is given by:
\begin{equation}\label{eq:P2}
    Tr[\rho^{(2)}\cdot\mathcal{O}^{(2)}_P] \approx (p_{T_1}^A+p_{T_1}^B)|\beta|^2+p_{reset}^Ap_{1}^A+p_{reset}^Bp_{1}^B,
\end{equation}
assuming $p_1^{A/B}\ll 1$. 
Both the signal and noise rates double as anticipated. For $N$ identical sensors operating independently, both the signal and noise rates scale linearly with $N$. The SNR and the signal significance can be derived from Eq.~\eqref{eq:SNR1} and \eqref{eq:sig1}
\begin{equation}
\label{eq:sig4}
    SNR \simeq \frac{p_{T_1}|\beta|^2}{(1-p_{T_1})\bar n_{r0}},\quad 
     \frac{S}{\sqrt{B}} \simeq  \frac{ N^{\frac{1}{2}} p_{T_1}|\beta|^2}{\sqrt{(1-p_{T_1})\bar n_{r0}}}.
\end{equation}

\subsection{Coherence-based detection}\label{sec:3a}
We now turn to the off-diagonal element of the density matrix, namely the coherence, and devise an observable $\mathcal{O}^{(2)}_C=|g_Ae_B\rangle\langle e_Ag_B|+|e_Ag_B\rangle\langle g_Ae_B|$, where the subscript $C$ indicates the coherence-channel, whose expectation value is given by
\begin{equation}\label{eq:nc}
    Tr[\rho^{(2)}\cdot\mathcal{O}^{(2)}_C]
    =2p_{T_2}^Ap_{T_2}^B|\beta|^2,
\end{equation}
which possesses comparable signal strength, but is completely noise free.

By implementing a unitary transformation $\mathcal{U}^{(2)}_C$ to diagonalize the operator $\mathcal{O}^{(2)}_C$, quantum coherence can be converted into population measurements that can be detected by quantum sensors. The unitary matrix $\mathcal{U}^{(2)}_C$ is not unique. As an example,
\begin{equation}\label{eq:uc}
    \mathcal{U}^{(2)}_C=\begin{pmatrix}
        1&0&0&0\\ 0&\frac{1}{\sqrt{2}}&\frac{1}{\sqrt{2}}&0\\ 0&-\frac{1}{\sqrt{2}}&\frac{1}{\sqrt{2}}&0\\ 0&0&0&1
    \end{pmatrix},
\end{equation}
where the matrix act on wave vectors with basis $\{|gg\rangle,|ge\rangle,|eg\rangle,|ee\rangle\}$, from top to bottom,
so that
\begin{equation}
\mathcal{O}'^{(2)}_P=\mathcal{U}^{(2)}_C\mathcal{O}^{(2)}_C\mathcal{U}^{(2)\dagger}_C=|g_Ae_B\rangle\langle g_Ae_B|-|e_Ag_B\rangle\langle e_Ag_B|.
\end{equation}
We now have
\begin{equation}
\label{eq:uro}
Tr[\rho^{(2)}\cdot\mathcal{O}^{(2)}_C]=Tr[\rho'^{(2)}\cdot\mathcal{O}'^{(2)}_P]=\bar{n}'_{g_Ae_B}-\bar{n}'_{e_Ag_B},
\end{equation}
where $\rho'^{(2)}=\mathcal{U}^{(2)}_C\rho^{(2)}_C\mathcal{U}^{(2)\dagger}_C$ is the transformed density matrix, $\bar{n}'_{i_Aj_B}=\langle i_Aj_B|\rho'^{(2)}|i_Aj_B\rangle$ represents the population of state $|i_Aj_B\rangle$ with $i,j$ equals $e$ or $g$. For the signal part, $\bar{n}'_{g_Ae_B} = [(p_{T_1}^A+p_{T_1}^B)/2+p_{T_2}^Ap_{T_2}^B]|\beta|^2$,  $\bar{n}'_{e_Ag_B} =[(p_{T_1}^A+p_{T_1}^B)/2-p_{T_2}^Ap_{T_2}^B]|\beta|^2$. For the noise part, $\bar{n}'_{g_Ae_B} = \bar{n}'_{e_Ag_B} =(p_{reset}^Ap_{1}^A+p_{reset}^Bp_{1}^B)/2$. Thus, for two detectors, after the DM signal undergoes a unitary transformation, we expect the signal to be concentrated in one detector, while the noise is evenly distributed between the two detectors. By subtracting the signals in the two detectors, we obtain the entire coherence signal and ideally zero noise.

For $N$ multiple quantum sensors, there are generally ${N(N-1)}/{2}$ coherence channels. Nevertheless, here we utilize only a subset of these coherence channels to define an observable. We divide the whole system into two parts, each containing $N/2$ sensors (assuming $N$ is even), and only consider the coherence between these two portions. As we will see in Section~\ref{sec:3b}, this definition results in reduced overall noise. The whole observable we are concerned is defined as the sum over the ${N^2}/{4}$ coherence channels between sensors $I(N\geq I>N/2)$ and $J(N/2\geq J\geq 1)$
\begin{equation}
    \mathcal{O}^{(N)}_{C}=\sum_{I>N/2\geq J} \mathcal{O}^{(2)}_{C_{IJ}} 
    = \sum_{I>N/2\geq J } |e_{I}^{(N)}\rangle\langle e_{J}^{(N)}| +h.c.,\label{obss}
\end{equation}
where $|e_{I}^{(N)}\rangle\equiv|g_1\dots e_I\dots g_{N}\rangle$ describes the system where the $I$-th sensor is in the excited state, while all other sensors remain in their ground states.
The signal is indicated by the expectation value
\begin{equation}\label{ocn}
    Tr[\rho^{(N)}\cdot\mathcal{O}^{(N)}_{C}]=\sum_{I>N/2\geq J}2 p_{T_2}^Ip_{T_2}^J|\beta|^2,
\end{equation}
which is noise-free and enhanced by $N^2 /2$, where $\rho^{(N)}=\mathcal{E}_1(\rho^{(1)}) \otimes \mathcal{E}_2(\rho^{(1)}) \otimes\cdots\otimes \mathcal{E}_{N}(\rho^{(1)})$ is the density matrix of $N$ sensors.

\begin{figure}[]
\centering
\includegraphics[width=0.45\textwidth]{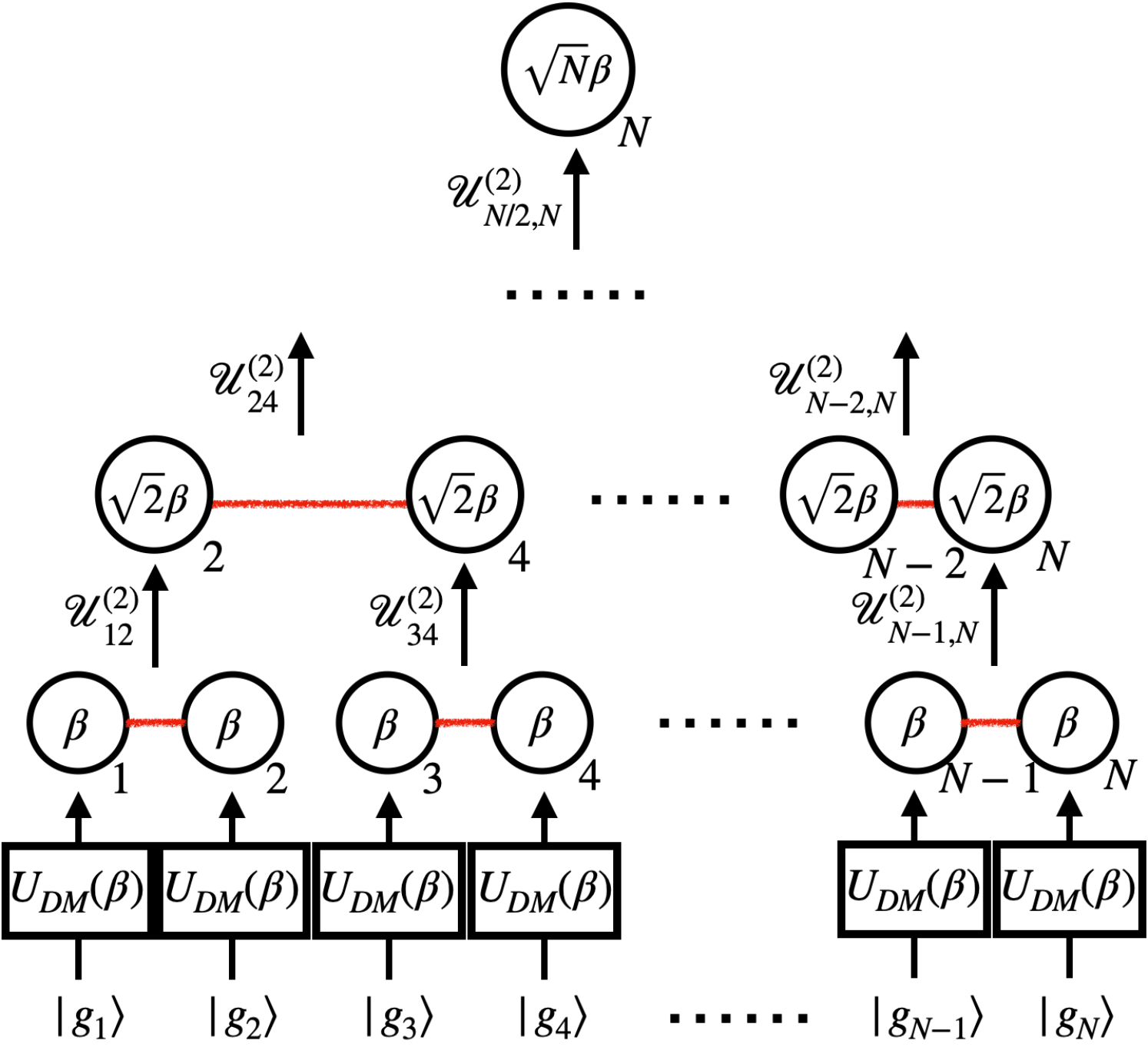}
\caption{The perfect binary tree structure of the coherence signal transfer route.
Firstly, we have $N$ sensors prepared in their ground states $|g_i\rangle$. The DM signal with strength $\beta$ then drives each sensor independently into state $|\beta\rangle$ as in Eq. \eqref{beta}.
Signal strengths are represented by the amplitudes in each circle while the numbers in the right bottom corner label the sequence of the sensor in an array.
The red link connecting two sensors $I$ and $J$ represents an implementation of $\mathcal{U}^{(2)}_{C_{IJ}}$, where the coherence signal between the two sensors is extracted and restored in the sensor $J$. And finally all of the coherence signals are saved in the last detector $N$ with amplitude amplified by $\sqrt{N}$. From the expectation of Eq. \eqref{obs} we conclude the signals are enhanced by $N^2/2$.
}
\label{fig:img}
\end{figure}

Quantum coherence can also be extracted in a similar way by executing a unitary transformation $\mathcal{U}^{(N)}_C$ to diagonalize $\mathcal{O}^{(N)}_C$, measuring the population and post-possessing the data. In general, $\mathcal{U}^{(N)}_C$ for $N=2^s$ can be decomposed by a series of $\mathcal{U}^{(2)}_{C_{IJ}}$'s in a divide-and-conquer manner which is illustrated in Fig~\ref{fig:img}
\begin{align}\nonumber
\mathcal{U}^{(N)}_C=&\mathcal{U}^{(2)}_{C_{N/2,N}}
(\mathcal{U}^{(2)}_{C_{N/4,N/2}}\mathcal{U}^{(2)}_{C_{3N/4,N}}) \\
&\cdots (\mathcal{U}^{(2)}_{C_{12}}\mathcal{U}^{(2)}_{C_{34}}\cdots\mathcal{U}^{(2)}_{C_{N-1,N}}),
\end{align}
which only needs $s=\log_2 N$ steps, and $\mathcal{U}^{(2)}_{C_{IJ}}$ is the unitary transformation of Eq. \eqref{eq:uc} acting on the Hilbert space $\mathcal{H}_I\otimes\mathcal{H}_J$ of sensors $I$ and $J$.
The transformed observable is
\begin{align}
\nonumber\mathcal{O}'^{(N)}_P=&\mathcal{U}^{(N)}_C\mathcal{O}^{(N)}_C\mathcal{U}^{(N)\dagger}_C\\
=&\frac{N}{2}\left(|e_{N}^{(N)}\rangle\langle e_{N}^{(N)}|-|e_{N/2}^{(N)}\rangle\langle e_{N/2}^{(N)}|\right).\label{obs}
\end{align}

Despite the perfect binary tree structure of the coherence signal transfer route as illustrated in Fig. \ref{fig:img}, $N$ does not necessarily be an order of 2. For a general $N$, $\mathcal{U}^{(N)}_C$ can always be decomposed into $(N-1)$ two-body unitary transformations and implemented within $\lceil\log_2 N\rceil$ steps, with the structure forming a complete binary tree. The signals obtained in the final sensor are noise-free and enhanced by a factor of $N^2/2$. See SM for more examples.

\subsection{Implementation and infidelities}
\label{sec:3b}
Due to hardware infidelities and environmental interactions, the application of $\mathcal{U}^{(N)}_C$ often introduces additional errors. 
Here, we use the depolarizing Pauli error model to mimic the hardware infidelity, which involves Pauli operations that can lead to errors with a certain probability.
However, when depolarizing errors act uniformly, as with the symmetric depolarizing channel, hardware infidelities do not contribute to background noise. Additional noise is introduced only when the depolarizing error acts unevenly on subsystem $A$ and $B$.
The expectation value in Eq.~\eqref{eq:uro} is then altered to be
\begin{align}\label{depo}
    \nonumber
    Tr&[\mathcal{D}_{AB}(\rho'^{(2)})\cdot\mathcal{O}'^{(2)}_{P}]=\\
    &\frac{\lambda_B-\lambda_A}{2}(1-p_{reset}^Ap_{1}^A-p_{reset}^Bp_{1}^B)+ \cdots  ,
\end{align}
where $\cdots$ represents the signal approximately equal to $2p_{T_2}^Ap_{T_2}^B|\beta|^2$, and $\mathcal{D}_{AB}(\rho)= \mathcal{D}_A(\mathcal{D}_B(\rho))$ denotes the combination of the depolarizing error on sensors $A$ and $B$ with different strength $\lambda_A$ and $\lambda_B$. The infidelity of the depolarizing channel can be expressed as $\bar{\mathcal{F}_{1}}\sim \lambda$. 

On the other hand, the interaction between the sensors and the environment can be modeled by the thermal relaxation and dephasing errors described in Eq. \eqref{eq:deco} with an interaction time of $\tau_g \ll T_1$. The corresponding expectation value can be calculated as 
\begin{align}
\label{eq:nde}
\nonumber  Tr[\mathcal{E}_{AB}(\rho'^{(2)})& \cdot\mathcal{O}'^{(2)}_{P}]= \frac{\epsilon_A + \epsilon_B}{2}(p_{1}^B-p_{1}^A)\\
&+\frac{\epsilon_B-\epsilon_A}{2}(p_{T_1}^Ap_{1}^A+p_{T_1}^Bp_{1}^B) + \cdots ,
\end{align}
where the infidelity $\bar{\mathcal{F}_{2}} = \epsilon_{A/B}=1-e^{-\tau_g/T_1^{A/B}}\simeq{\tau_g}/{T_1^{A/B}}$, and $p_1^{A/B} \sim \bar{n}_r$, 
$\mathcal{E}_{AB}(\rho)= \mathcal{E}_A(\mathcal{E}_B(\rho))$ is the combination of the error channels $\mathcal{E}_A$ and $\mathcal{E}_B$.
Therefore, the background noise added by the thermal decoherence channel is of the order of $O(\bar n_r\tau_g/T_1)$.

Now, we can estimate the potential sensitivity when considering hardware infidelities in various situations. In coherence-based detection, the signal scales as $N^2p_{T_2}^2 |\beta|^2/2$, 
while background noise is proportional to $N\bar{\mathcal{F}}/2$, which is added only in the last implementation of $\mathcal{U}^{(2)}_{C_{IJ}}$. This is because the errors added in the previous steps can all be considered as incoherent noises between the two portions that we divide the system into, and are canceled in the last step following the same story as the two-sensor case.
If there are significant uneven depolarizing errors, they will dominate the background noise. we thus have a conservative estimation of the SNR and the signal significance for the coherence-based detection.
\begin{equation}\label{eq:sig2}
    SNR\simeq N\frac{p_{T_2}^2 |\beta|^2}{\bar{\mathcal{F}}},\quad 
    \frac{S}{\sqrt{B}}\simeq N^{\frac{3}{2}}\frac{p_{T_2}^2|\beta|^2}{\sqrt{2 \bar{\mathcal{F}}}}.
\end{equation}
In some special cases where only the symmetric depolarizing channel is present, the error budget is dominated by the decoherence error channel, which is limited only by the operation time $\tau_g$. From Eq.~\eqref{eq:nde}, we see that the background noise is further suppressed by a factor of $\bar n_r$. In such cases, we have
\begin{equation}
\label{eq:sig3}
    SNR\simeq N\frac{p_{T_2}^2 |\beta|^2}{\bar n_r\bar{\mathcal{F}}},\quad 
    \frac{S}{\sqrt{B}}\simeq N^{\frac32}\frac{p_{T_2}^2 |\beta|^2}{\sqrt{2 \bar n_r\bar{\mathcal{F}}}}.
\end{equation}

\begin{figure}[]
\centering
\includegraphics[width=0.45\textwidth]{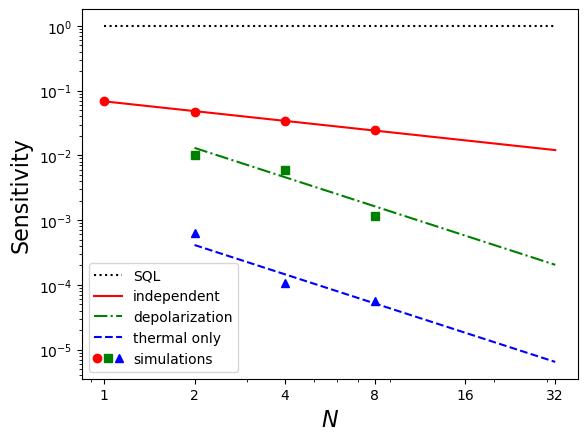}
\caption{
Detection sensitivities with $1\sigma$ significance ($S/\sqrt{B}=1$)
as a function of the number of sensors $N$ for cavities as examples (transmons and trapped ions would have lower sensitivities due to their higher $\bar n_{r0}$ and $\bar{\mathcal{F}}$).
The black dotted line shows the SQL with $\bar n=1$. The solid line represents sensitivities with $N$ independent sensors without coherence detection given by Eq. \eqref{eq:sig4}. The dotted-dashed and dashed lines represent sensitivities reached by coherence based detection with proper normalization where the noise is dominated by gate infidelities (Eq. \eqref{eq:sig2}) or purely decoherence errors (Eq. \eqref{eq:sig3}), respectively. The dots, squares and triangles represent numerical simulation results for $N$ sensors with independent detection and coherent-based detection with noises mainly contributed by hardware infidelities or thermal relaxations, respectively.
}
\label{fig1}
\end{figure}

The thermal excitation at very low temperatures typically saturates to a small number $\bar n_{r0}$. The two-qubit gate infidelity, $\bar{\mathcal{F}}$, encompasses both hardware infidelity from the depolarizing channel and the decoherence channel. The parameters of different types of detectors are summarized in Table~\ref{tb:para}. We plot the projected detection sensitivities for cavities in Fig~\ref{fig1} for taking the signal significance $S/\sqrt{B}=1$ using Eqs. \eqref{eq:sig4}, \eqref{eq:sig2} and \eqref{eq:sig3} as a conservative estimation.
We observe that the sensitivities are well below the SQL $\bar n=1$ due to the QND measurement merit of the detectors. For population detection using independent sensors, the sensitivity scales as $N^{-1/2}$, while for coherence-based detection using correlated sensors, the sensitivity scales as $N^{-3/2}$.

\begin{table}[]
	\centering
	\begin{tabular}{c|c|c}
		\hline
		Sensor type & $\bar n_{r0}$ & Two-qubit gate infidelity $\bar{\mathcal{F}}$ \\
		\hline
		Cavity & $10^{-3}$~\cite{ni2023} & $10^{-4}$~\cite{lu2023high} \\
		\hline
		Transmon & $10^{-3}$~\cite{jin2015thermal,serniak2018hot} & $10^{-3}$~\cite{arute2019quantum,sung2021} \\ 
		\hline
		Ions & $10^{-2}$~\cite{RevModPhys.87.1419}& $10^{-3}$~\cite{PhysRevLett.117.060504, PhysRevLett.117.060505} \\
		\hline
	\end{tabular}
	\caption{Parameters for different types of detectors.}
	\label{tb:para}
\end{table}

We also use QISKIT~\cite{qiskit2024} to simulate the quantum circuit shown in Fig.~\ref{fig:img}. Initially, we create a quantum register with $N$ qubits, each initialized in their ground state $|g\rangle$. The influence of DM is simulated by applying a set of single-qubit gates $U_{DM}(\beta) = \exp[\beta |e\rangle\langle g| - \beta^* |g\rangle\langle e|]$ to each qubit independently. These gates include decoherence errors parameterized by the integration time $\tau$, thermal relaxation times $T_1$, dephasing times $T_2$, and the populations of the excited state at equilibrium $p_1$. The ratios $T_1/\tau$ and $T_2/\tau$ are randomly drawn from a uniform distribution between $[0.8, 1.2]$, and $p_1=\bar n_{r0}$ is set at $10^{-3}$. The two-qubit unitary transformation $\mathcal{U}^{(2)}_{C_{IJ}}$ is simulated by a two-qubit gate $U^{(2)} = \exp[\frac{\pi}{4} (|ge\rangle\langle eg| - |eg\rangle\langle ge|)]$ with a gate time $\tau_g = 10^{-4}\tau$. During this period, hardware infidelities are simulated by a depolarizing error that affects the two qubits unevenly, with strength $\lambda$ drawn from a uniform distribution between $[0.5, 1.5] \times 10^{-4}$. Thermal relaxation errors are also simulated with the given parameters. The circuit is run repeatedly enough times to suppress any statistical errors. Finally, we count the population of excitations for the $N$-th and $N/2$-th sensors and combine them according to Eq.~\eqref{obs}.

We characterize the background noise by extrapolating $|\beta|^2\rightarrow 0$. And we vary the signal strength $|\beta|^2$ until the significance $S/\sqrt{B}>1$ to obtain the sensitivity. The final results are shown in Fig. \ref{fig1}. The simulation results show that our coherence based detection method outperforms $N$ independent sensors and the sensitivity scales as $N^{-3/2}$, as expected.

\section{Outlook}\label{sec:out}

In this work, we propose a novel DM detection scheme that utilizes the quantum coherence channels among sensors. This approach surpasses conventional population-based sensor measurements due to its immunity to incoherent background noise. Disabling interactions between sensors during the detection process ensures that noise from excitation events remains uncorrelated across sensors. Conversely, quantum sensors within their coherence length and time are coherently excited by bosonic DM fields, thus providing a inherently noise-free channel. In a network comprising $N$ sensors, there exist a total of $O(N^2)$ coherence channels. The noise-free coherence signals are extracted by initially applying a unitary transformation and subsequently performing a population measurement in the transformed basis. This transformation is achieved through $O(N)$ two-body operations within a duration of $O(\tau_g\log_2 N)$. Consequently, the resulting signal strength scales as $O(N^2)$, with only an additional $O(N)$ noise, resulting in a signal-to-noise ratio that scales linearly with $N$. The DM detection sensitivity is primarily limited by the infidelities of the basis-transformation operations rather than incoherent background noise. Additionally, the logarithmic scaling of the operation time mitigates constraints associated with scalability.

In addition to employing vacuum states as the initial probe states, a range of nonclassical states—including Fock states~\cite{Agrawal:2023umy, deng2023}, Greenberger–Horne–Zeilinger states~\cite{Chen:2023swh, Ito:2023zhp}, squeezed states~\cite{caves1981quantum, lawrie2019quantum, eickbusch2022fast}, and Schrödinger cat states~\cite{ourjoumtsev2007, vlastakis2013, ofek2016extending,milul2023,pan2024}—emerge as promising candidates for enhancing the capabilities of DM detection. Further investigation into the integration of these exotic quantum states within our coherence-based method hold the potential to substantially improve the DM detection sensitivity.

Furthermore, by leveraging the alternating strong electromagnetic fields confined in superconducting cavities~\cite{Berlin:2019ahk,Berlin:2020vrk} or the Nb-based Josephson junctions and transmon qubits capable of tolerating strong magnetic fields~\cite{Chen:2024aya,krause2022magnetic,rokhinson2012fractional}, our approach facilitates advanced measurements of axions and ultra-high-frequency gravitation waves~\cite{Aggarwal:2020olq, Berlin:2021txa} under these strong magnetic field conditions.

\begin{acknowledgments}
This work is supported by the National Key Research and Development Program of China under Grant No. 2020YFC2201501 and 2021YFC2203004. J.S. is supported by Peking University under startup Grant No. 7101302974 and the National Natural Science Foundation of China under Grants No. 12025507, No.12150015; and is supported by the Key Research Program of Frontier Science of the Chinese Academy of Sciences (CAS) under Grants No. ZDBS-LY-7003. Y.X. is supported by the National Natural Science Foundation of China (Grants No. 12422416, No. 12274198), the Shenzhen Science and Technology Program (Grant No. RCYX20210706092103021), the Guangdong Basic and Applied Basic Research Foundation (Grant No. 2024B1515020013), and the Innovation Program for Quantum Science and Technology (Grant No. ZD0301703).
\end{acknowledgments}

\hspace{5mm}
\bibliography{refs}

\begin{thebibliography}{75}%
\makeatletter
\providecommand \@ifxundefined [1]{%
 \@ifx{#1\undefined}
}%
\providecommand \@ifnum [1]{%
 \ifnum #1\expandafter \@firstoftwo
 \else \expandafter \@secondoftwo
 \fi
}%
\providecommand \@ifx [1]{%
 \ifx #1\expandafter \@firstoftwo
 \else \expandafter \@secondoftwo
 \fi
}%
\providecommand \natexlab [1]{#1}%
\providecommand \enquote  [1]{``#1''}%
\providecommand \bibnamefont  [1]{#1}%
\providecommand \bibfnamefont [1]{#1}%
\providecommand \citenamefont [1]{#1}%
\providecommand \href@noop [0]{\@secondoftwo}%
\providecommand \href [0]{\begingroup \@sanitize@url \@href}%
\providecommand \@href[1]{\@@startlink{#1}\@@href}%
\providecommand \@@href[1]{\endgroup#1\@@endlink}%
\providecommand \@sanitize@url [0]{\catcode `\\12\catcode `\$12\catcode
  `\&12\catcode `\#12\catcode `\^12\catcode `\_12\catcode `\%12\relax}%
\providecommand \@@startlink[1]{}%
\providecommand \@@endlink[0]{}%
\providecommand \url  [0]{\begingroup\@sanitize@url \@url }%
\providecommand \@url [1]{\endgroup\@href {#1}{\urlprefix }}%
\providecommand \urlprefix  [0]{URL }%
\providecommand \Eprint [0]{\href }%
\providecommand \doibase [0]{http://dx.doi.org/}%
\providecommand \selectlanguage [0]{\@gobble}%
\providecommand \bibinfo  [0]{\@secondoftwo}%
\providecommand \bibfield  [0]{\@secondoftwo}%
\providecommand \translation [1]{[#1]}%
\providecommand \BibitemOpen [0]{}%
\providecommand \bibitemStop [0]{}%
\providecommand \bibitemNoStop [0]{.\EOS\space}%
\providecommand \EOS [0]{\spacefactor3000\relax}%
\providecommand \BibitemShut  [1]{\csname bibitem#1\endcsname}%
\let\auto@bib@innerbib\@empty
\bibitem [{\citenamefont {Gianfranco}(2010)}]{bertone_2010}%
  \BibitemOpen
  \bibfield  {author} {\bibinfo {author} {\bibfnamefont {B.}~\bibnamefont
  {Gianfranco}},\ }\href {\doibase 10.1017/CBO9780511770739} {\emph {\bibinfo
  {title} {Particle Dark Matter: Observations, Models and Searches}}}\
  (\bibinfo  {publisher} {Cambridge University Press},\ \bibinfo {year}
  {2010})\BibitemShut {NoStop}%
\bibitem [{\citenamefont {Salucci}(2019)}]{Salucci_2019}%
  \BibitemOpen
  \bibfield  {author} {\bibinfo {author} {\bibfnamefont {Paolo}\ \bibnamefont
  {Salucci}},\ }\bibfield  {title} {\enquote {\bibinfo {title} {The
  distribution of dark matter in galaxies},}\ }\href {\doibase
  10.1007/s00159-018-0113-1} {\bibfield  {journal} {\bibinfo  {journal} {The
  Astronomy and Astrophysics Review}\ }\textbf {\bibinfo {volume} {27}},\
  \bibinfo {pages} {2} (\bibinfo {year} {2019})}\BibitemShut {NoStop}%
\bibitem [{\citenamefont {Sofue}\ and\ \citenamefont
  {Rubin}(2001)}]{doi:10.1146/annurev.astro.39.1.137}%
  \BibitemOpen
  \bibfield  {author} {\bibinfo {author} {\bibfnamefont {Yoshiaki}\
  \bibnamefont {Sofue}}\ and\ \bibinfo {author} {\bibfnamefont {Vera}\
  \bibnamefont {Rubin}},\ }\bibfield  {title} {\enquote {\bibinfo {title}
  {Rotation curves of spiral galaxies},}\ }\href {\doibase
  10.1146/annurev.astro.39.1.137} {\bibfield  {journal} {\bibinfo  {journal}
  {Annu. Rev. Astron. Astrophys.}\ }\textbf {\bibinfo {volume} {39}},\ \bibinfo
  {pages} {137--174} (\bibinfo {year} {2001})}\BibitemShut {NoStop}%
\bibitem [{\citenamefont {Massey}\ \emph {et~al.}(2010)\citenamefont {Massey},
  \citenamefont {Kitching},\ and\ \citenamefont {Richard}}]{Massey_2010}%
  \BibitemOpen
  \bibfield  {author} {\bibinfo {author} {\bibfnamefont {Richard}\ \bibnamefont
  {Massey}}, \bibinfo {author} {\bibfnamefont {Thomas}\ \bibnamefont
  {Kitching}}, \ and\ \bibinfo {author} {\bibfnamefont {Johan}\ \bibnamefont
  {Richard}},\ }\bibfield  {title} {\enquote {\bibinfo {title} {The dark matter
  of gravitational lensing},}\ }\href {\doibase 10.1088/0034-4885/73/8/086901}
  {\bibfield  {journal} {\bibinfo  {journal} {Rep. Prog. Phys.}\ }\textbf
  {\bibinfo {volume} {73}},\ \bibinfo {pages} {086901} (\bibinfo {year}
  {2010})}\BibitemShut {NoStop}%
\bibitem [{\citenamefont {Markevitch}\ \emph {et~al.}(2004)\citenamefont
  {Markevitch}, \citenamefont {Gonzalez}, \citenamefont {Clowe}, \citenamefont
  {Vikhlinin}, \citenamefont {Forman}, \citenamefont {Jones}, \citenamefont
  {Murray},\ and\ \citenamefont {Tucker}}]{Markevitch_2004}%
  \BibitemOpen
  \bibfield  {author} {\bibinfo {author} {\bibfnamefont {M.}~\bibnamefont
  {Markevitch}}, \bibinfo {author} {\bibfnamefont {A.~H.}\ \bibnamefont
  {Gonzalez}}, \bibinfo {author} {\bibfnamefont {D.}~\bibnamefont {Clowe}},
  \bibinfo {author} {\bibfnamefont {A.}~\bibnamefont {Vikhlinin}}, \bibinfo
  {author} {\bibfnamefont {W.}~\bibnamefont {Forman}}, \bibinfo {author}
  {\bibfnamefont {C.}~\bibnamefont {Jones}}, \bibinfo {author} {\bibfnamefont
  {S.}~\bibnamefont {Murray}}, \ and\ \bibinfo {author} {\bibfnamefont
  {W.}~\bibnamefont {Tucker}},\ }\bibfield  {title} {\enquote {\bibinfo {title}
  {Direct constraints on the dark matter self-interaction cross section from
  the merging galaxy cluster 1e 0657–56},}\ }\href {\doibase 10.1086/383178}
  {\bibfield  {journal} {\bibinfo  {journal} {The Astrophysical Journal}\
  }\textbf {\bibinfo {volume} {606}},\ \bibinfo {pages} {819} (\bibinfo {year}
  {2004})}\BibitemShut {NoStop}%
\bibitem [{\citenamefont {Arias}\ \emph {et~al.}(2012)\citenamefont {Arias},
  \citenamefont {Cadamuro}, \citenamefont {Goodsell}, \citenamefont {Jaeckel},
  \citenamefont {Redondo},\ and\ \citenamefont {Ringwald}}]{Arias:2012az}%
  \BibitemOpen
  \bibfield  {author} {\bibinfo {author} {\bibfnamefont {Paola}\ \bibnamefont
  {Arias}}, \bibinfo {author} {\bibfnamefont {Davide}\ \bibnamefont
  {Cadamuro}}, \bibinfo {author} {\bibfnamefont {Mark}\ \bibnamefont
  {Goodsell}}, \bibinfo {author} {\bibfnamefont {Joerg}\ \bibnamefont
  {Jaeckel}}, \bibinfo {author} {\bibfnamefont {Javier}\ \bibnamefont
  {Redondo}}, \ and\ \bibinfo {author} {\bibfnamefont {Andreas}\ \bibnamefont
  {Ringwald}},\ }\bibfield  {title} {\enquote {\bibinfo {title} {{WISPy Cold
  Dark Matter}},}\ }\href {\doibase 10.1088/1475-7516/2012/06/013} {\bibfield
  {journal} {\bibinfo  {journal} {JCAP}\ }\textbf {\bibinfo {volume} {06}},\
  \bibinfo {pages} {013} (\bibinfo {year} {2012})},\ \Eprint
  {http://arxiv.org/abs/1201.5902} {arXiv:1201.5902 [hep-ph]} \BibitemShut
  {NoStop}%
\bibitem [{\citenamefont {Jackson~Kimball}\ and\ \citenamefont
  {Van~Bibber}(2023)}]{jackson2023search}%
  \BibitemOpen
  \bibfield  {author} {\bibinfo {author} {\bibfnamefont {Derek~F}\ \bibnamefont
  {Jackson~Kimball}}\ and\ \bibinfo {author} {\bibfnamefont {Karl}\
  \bibnamefont {Van~Bibber}},\ }\href@noop {} {\emph {\bibinfo {title} {The
  search for ultralight bosonic dark matter}}}\ (\bibinfo  {publisher}
  {Springer Nature},\ \bibinfo {year} {2023})\BibitemShut {NoStop}%
\bibitem [{\citenamefont {Preskill}\ \emph {et~al.}(1983)\citenamefont
  {Preskill}, \citenamefont {Wise},\ and\ \citenamefont
  {Wilczek}}]{Preskill:1982cy}%
  \BibitemOpen
  \bibfield  {author} {\bibinfo {author} {\bibfnamefont {John}\ \bibnamefont
  {Preskill}}, \bibinfo {author} {\bibfnamefont {Mark~B.}\ \bibnamefont
  {Wise}}, \ and\ \bibinfo {author} {\bibfnamefont {Frank}\ \bibnamefont
  {Wilczek}},\ }\bibfield  {title} {\enquote {\bibinfo {title} {{Cosmology of
  the Invisible Axion}},}\ }\href {\doibase 10.1016/0370-2693(83)90637-8}
  {\bibfield  {journal} {\bibinfo  {journal} {Phys. Lett. B}\ }\textbf
  {\bibinfo {volume} {120}},\ \bibinfo {pages} {127--132} (\bibinfo {year}
  {1983})}\BibitemShut {NoStop}%
\bibitem [{\citenamefont {Abbott}\ and\ \citenamefont
  {Sikivie}(1983)}]{Abbott:1982af}%
  \BibitemOpen
  \bibfield  {author} {\bibinfo {author} {\bibfnamefont {L.~F.}\ \bibnamefont
  {Abbott}}\ and\ \bibinfo {author} {\bibfnamefont {P.}~\bibnamefont
  {Sikivie}},\ }\bibfield  {title} {\enquote {\bibinfo {title} {{A Cosmological
  Bound on the Invisible Axion}},}\ }\href {\doibase
  10.1016/0370-2693(83)90638-X} {\bibfield  {journal} {\bibinfo  {journal}
  {Phys. Lett. B}\ }\textbf {\bibinfo {volume} {120}},\ \bibinfo {pages}
  {133--136} (\bibinfo {year} {1983})}\BibitemShut {NoStop}%
\bibitem [{\citenamefont {Dine}\ and\ \citenamefont
  {Fischler}(1983)}]{Dine:1982ah}%
  \BibitemOpen
  \bibfield  {author} {\bibinfo {author} {\bibfnamefont {Michael}\ \bibnamefont
  {Dine}}\ and\ \bibinfo {author} {\bibfnamefont {Willy}\ \bibnamefont
  {Fischler}},\ }\bibfield  {title} {\enquote {\bibinfo {title} {{The Not So
  Harmless Axion}},}\ }\href {\doibase 10.1016/0370-2693(83)90639-1} {\bibfield
   {journal} {\bibinfo  {journal} {Phys. Lett. B}\ }\textbf {\bibinfo {volume}
  {120}},\ \bibinfo {pages} {137--141} (\bibinfo {year} {1983})}\BibitemShut
  {NoStop}%
\bibitem [{\citenamefont {Graham}\ \emph {et~al.}(2015)\citenamefont {Graham},
  \citenamefont {Kaplan},\ and\ \citenamefont {Rajendran}}]{Graham:2015cka}%
  \BibitemOpen
  \bibfield  {author} {\bibinfo {author} {\bibfnamefont {Peter~W.}\
  \bibnamefont {Graham}}, \bibinfo {author} {\bibfnamefont {David~E.}\
  \bibnamefont {Kaplan}}, \ and\ \bibinfo {author} {\bibfnamefont {Surjeet}\
  \bibnamefont {Rajendran}},\ }\bibfield  {title} {\enquote {\bibinfo {title}
  {{Cosmological Relaxation of the Electroweak Scale}},}\ }\href {\doibase
  10.1103/PhysRevLett.115.221801} {\bibfield  {journal} {\bibinfo  {journal}
  {Phys. Rev. Lett.}\ }\textbf {\bibinfo {volume} {115}},\ \bibinfo {pages}
  {221801} (\bibinfo {year} {2015})},\ \Eprint
  {http://arxiv.org/abs/1504.07551} {arXiv:1504.07551 [hep-ph]} \BibitemShut
  {NoStop}%
\bibitem [{\citenamefont {Co}\ \emph {et~al.}(2021)\citenamefont {Co},
  \citenamefont {Hall},\ and\ \citenamefont {Harigaya}}]{Co:2020xlh}%
  \BibitemOpen
  \bibfield  {author} {\bibinfo {author} {\bibfnamefont {Raymond~T.}\
  \bibnamefont {Co}}, \bibinfo {author} {\bibfnamefont {Lawrence~J.}\
  \bibnamefont {Hall}}, \ and\ \bibinfo {author} {\bibfnamefont {Keisuke}\
  \bibnamefont {Harigaya}},\ }\bibfield  {title} {\enquote {\bibinfo {title}
  {{Predictions for Axion Couplings from ALP Cogenesis}},}\ }\href {\doibase
  10.1007/JHEP01(2021)172} {\bibfield  {journal} {\bibinfo  {journal} {JHEP}\
  }\textbf {\bibinfo {volume} {01}},\ \bibinfo {pages} {172} (\bibinfo {year}
  {2021})},\ \Eprint {http://arxiv.org/abs/2006.04809} {arXiv:2006.04809
  [hep-ph]} \BibitemShut {NoStop}%
\bibitem [{\citenamefont {Holdom}(1986{\natexlab{a}})}]{holdom1986two}%
  \BibitemOpen
  \bibfield  {author} {\bibinfo {author} {\bibfnamefont {Bob}\ \bibnamefont
  {Holdom}},\ }\bibfield  {title} {\enquote {\bibinfo {title} {Two u (1)'s and
  $\epsilon$ charge shifts},}\ }\href@noop {} {\bibfield  {journal} {\bibinfo
  {journal} {Phys. Lett. B}\ }\textbf {\bibinfo {volume} {166}},\ \bibinfo
  {pages} {196--198} (\bibinfo {year} {1986}{\natexlab{a}})}\BibitemShut
  {NoStop}%
\bibitem [{\citenamefont {Nelson}\ and\ \citenamefont
  {Scholtz}(2011)}]{Nelson:2011sf}%
  \BibitemOpen
  \bibfield  {author} {\bibinfo {author} {\bibfnamefont {Ann~E.}\ \bibnamefont
  {Nelson}}\ and\ \bibinfo {author} {\bibfnamefont {Jakub}\ \bibnamefont
  {Scholtz}},\ }\bibfield  {title} {\enquote {\bibinfo {title} {{Dark Light,
  Dark Matter and the Misalignment Mechanism}},}\ }\href {\doibase
  10.1103/PhysRevD.84.103501} {\bibfield  {journal} {\bibinfo  {journal} {Phys.
  Rev. D}\ }\textbf {\bibinfo {volume} {84}},\ \bibinfo {pages} {103501}
  (\bibinfo {year} {2011})},\ \Eprint {http://arxiv.org/abs/1105.2812}
  {arXiv:1105.2812 [hep-ph]} \BibitemShut {NoStop}%
\bibitem [{\citenamefont {Graham}\ \emph {et~al.}(2016)\citenamefont {Graham},
  \citenamefont {Mardon},\ and\ \citenamefont {Rajendran}}]{Graham:2015rva}%
  \BibitemOpen
  \bibfield  {author} {\bibinfo {author} {\bibfnamefont {Peter~W.}\
  \bibnamefont {Graham}}, \bibinfo {author} {\bibfnamefont {Jeremy}\
  \bibnamefont {Mardon}}, \ and\ \bibinfo {author} {\bibfnamefont {Surjeet}\
  \bibnamefont {Rajendran}},\ }\bibfield  {title} {\enquote {\bibinfo {title}
  {{Vector Dark Matter from Inflationary Fluctuations}},}\ }\href {\doibase
  10.1103/PhysRevD.93.103520} {\bibfield  {journal} {\bibinfo  {journal} {Phys.
  Rev. D}\ }\textbf {\bibinfo {volume} {93}},\ \bibinfo {pages} {103520}
  (\bibinfo {year} {2016})},\ \Eprint {http://arxiv.org/abs/1504.02102}
  {arXiv:1504.02102 [hep-ph]} \BibitemShut {NoStop}%
\bibitem [{\citenamefont {Cvetic}\ and\ \citenamefont
  {Langacker}(1996)}]{Cvetic:1995rj}%
  \BibitemOpen
  \bibfield  {author} {\bibinfo {author} {\bibfnamefont {Mirjam}\ \bibnamefont
  {Cvetic}}\ and\ \bibinfo {author} {\bibfnamefont {Paul}\ \bibnamefont
  {Langacker}},\ }\bibfield  {title} {\enquote {\bibinfo {title} {{Implications
  of Abelian extended gauge structures from string models}},}\ }\href {\doibase
  10.1103/PhysRevD.54.3570} {\bibfield  {journal} {\bibinfo  {journal} {Phys.
  Rev. D}\ }\textbf {\bibinfo {volume} {54}},\ \bibinfo {pages} {3570--3579}
  (\bibinfo {year} {1996})},\ \Eprint {http://arxiv.org/abs/hep-ph/9511378}
  {arXiv:hep-ph/9511378} \BibitemShut {NoStop}%
\bibitem [{\citenamefont {Svrcek}\ and\ \citenamefont
  {Witten}(2006)}]{Svrcek:2006yi}%
  \BibitemOpen
  \bibfield  {author} {\bibinfo {author} {\bibfnamefont {Peter}\ \bibnamefont
  {Svrcek}}\ and\ \bibinfo {author} {\bibfnamefont {Edward}\ \bibnamefont
  {Witten}},\ }\bibfield  {title} {\enquote {\bibinfo {title} {{Axions In
  String Theory}},}\ }\href {\doibase 10.1088/1126-6708/2006/06/051} {\bibfield
   {journal} {\bibinfo  {journal} {JHEP}\ }\textbf {\bibinfo {volume} {06}},\
  \bibinfo {pages} {051} (\bibinfo {year} {2006})},\ \Eprint
  {http://arxiv.org/abs/hep-th/0605206} {arXiv:hep-th/0605206} \BibitemShut
  {NoStop}%
\bibitem [{\citenamefont {Abel}\ \emph {et~al.}(2008)\citenamefont {Abel},
  \citenamefont {Goodsell}, \citenamefont {Jaeckel}, \citenamefont {Khoze},\
  and\ \citenamefont {Ringwald}}]{Abel:2008ai}%
  \BibitemOpen
  \bibfield  {author} {\bibinfo {author} {\bibfnamefont {S.~A.}\ \bibnamefont
  {Abel}}, \bibinfo {author} {\bibfnamefont {M.~D.}\ \bibnamefont {Goodsell}},
  \bibinfo {author} {\bibfnamefont {J.}~\bibnamefont {Jaeckel}}, \bibinfo
  {author} {\bibfnamefont {V.~V.}\ \bibnamefont {Khoze}}, \ and\ \bibinfo
  {author} {\bibfnamefont {A.}~\bibnamefont {Ringwald}},\ }\bibfield  {title}
  {\enquote {\bibinfo {title} {{Kinetic Mixing of the Photon with Hidden U(1)s
  in String Phenomenology}},}\ }\href {\doibase 10.1088/1126-6708/2008/07/124}
  {\bibfield  {journal} {\bibinfo  {journal} {JHEP}\ }\textbf {\bibinfo
  {volume} {07}},\ \bibinfo {pages} {124} (\bibinfo {year} {2008})},\ \Eprint
  {http://arxiv.org/abs/0803.1449} {arXiv:0803.1449 [hep-ph]} \BibitemShut
  {NoStop}%
\bibitem [{\citenamefont {Arvanitaki}\ \emph {et~al.}(2010)\citenamefont
  {Arvanitaki}, \citenamefont {Dimopoulos}, \citenamefont {Dubovsky},
  \citenamefont {Kaloper},\ and\ \citenamefont
  {March-Russell}}]{Arvanitaki:2009fg}%
  \BibitemOpen
  \bibfield  {author} {\bibinfo {author} {\bibfnamefont {Asimina}\ \bibnamefont
  {Arvanitaki}}, \bibinfo {author} {\bibfnamefont {Savas}\ \bibnamefont
  {Dimopoulos}}, \bibinfo {author} {\bibfnamefont {Sergei}\ \bibnamefont
  {Dubovsky}}, \bibinfo {author} {\bibfnamefont {Nemanja}\ \bibnamefont
  {Kaloper}}, \ and\ \bibinfo {author} {\bibfnamefont {John}\ \bibnamefont
  {March-Russell}},\ }\bibfield  {title} {\enquote {\bibinfo {title} {{String
  Axiverse}},}\ }\href {\doibase 10.1103/PhysRevD.81.123530} {\bibfield
  {journal} {\bibinfo  {journal} {Phys. Rev. D}\ }\textbf {\bibinfo {volume}
  {81}},\ \bibinfo {pages} {123530} (\bibinfo {year} {2010})},\ \Eprint
  {http://arxiv.org/abs/0905.4720} {arXiv:0905.4720 [hep-th]} \BibitemShut
  {NoStop}%
\bibitem [{\citenamefont {Goodsell}\ \emph {et~al.}(2009)\citenamefont
  {Goodsell}, \citenamefont {Jaeckel}, \citenamefont {Redondo},\ and\
  \citenamefont {Ringwald}}]{Goodsell:2009xc}%
  \BibitemOpen
  \bibfield  {author} {\bibinfo {author} {\bibfnamefont {Mark}\ \bibnamefont
  {Goodsell}}, \bibinfo {author} {\bibfnamefont {Joerg}\ \bibnamefont
  {Jaeckel}}, \bibinfo {author} {\bibfnamefont {Javier}\ \bibnamefont
  {Redondo}}, \ and\ \bibinfo {author} {\bibfnamefont {Andreas}\ \bibnamefont
  {Ringwald}},\ }\bibfield  {title} {\enquote {\bibinfo {title} {{Naturally
  Light Hidden Photons in LARGE Volume String Compactifications}},}\ }\href
  {\doibase 10.1088/1126-6708/2009/11/027} {\bibfield  {journal} {\bibinfo
  {journal} {JHEP}\ }\textbf {\bibinfo {volume} {11}},\ \bibinfo {pages} {027}
  (\bibinfo {year} {2009})},\ \Eprint {http://arxiv.org/abs/0909.0515}
  {arXiv:0909.0515 [hep-ph]} \BibitemShut {NoStop}%
\bibitem [{\citenamefont {Lin}\ \emph {et~al.}(2018)\citenamefont {Lin},
  \citenamefont {Schive}, \citenamefont {Wong},\ and\ \citenamefont
  {Chiueh}}]{Lin:2018whl}%
  \BibitemOpen
  \bibfield  {author} {\bibinfo {author} {\bibfnamefont {Shan-Chang}\
  \bibnamefont {Lin}}, \bibinfo {author} {\bibfnamefont {Hsi-Yu}\ \bibnamefont
  {Schive}}, \bibinfo {author} {\bibfnamefont {Shing-Kwong}\ \bibnamefont
  {Wong}}, \ and\ \bibinfo {author} {\bibfnamefont {Tzihong}\ \bibnamefont
  {Chiueh}},\ }\bibfield  {title} {\enquote {\bibinfo {title} {{Self-consistent
  construction of virialized wave dark matter halos}},}\ }\href {\doibase
  10.1103/PhysRevD.97.103523} {\bibfield  {journal} {\bibinfo  {journal} {Phys.
  Rev. D}\ }\textbf {\bibinfo {volume} {97}},\ \bibinfo {pages} {103523}
  (\bibinfo {year} {2018})},\ \Eprint {http://arxiv.org/abs/1801.02320}
  {arXiv:1801.02320 [astro-ph.CO]} \BibitemShut {NoStop}%
\bibitem [{\citenamefont {Centers}\ \emph {et~al.}(2021)\citenamefont {Centers}
  \emph {et~al.}}]{Centers:2019dyn}%
  \BibitemOpen
  \bibfield  {author} {\bibinfo {author} {\bibfnamefont {Gary~P.}\ \bibnamefont
  {Centers}} \emph {et~al.},\ }\bibfield  {title} {\enquote {\bibinfo {title}
  {{Stochastic fluctuations of bosonic dark matter}},}\ }\href {\doibase
  10.1038/s41467-021-27632-7} {\bibfield  {journal} {\bibinfo  {journal} {Nat.
  Commun.}\ }\textbf {\bibinfo {volume} {12}},\ \bibinfo {pages} {7321}
  (\bibinfo {year} {2021})},\ \Eprint {http://arxiv.org/abs/1905.13650}
  {arXiv:1905.13650 [astro-ph.CO]} \BibitemShut {NoStop}%
\bibitem [{\citenamefont {Holdom}(1986{\natexlab{b}})}]{Holdom:1986ag}%
  \BibitemOpen
  \bibfield  {author} {\bibinfo {author} {\bibfnamefont {Bob}\ \bibnamefont
  {Holdom}},\ }\bibfield  {title} {\enquote {\bibinfo {title} {{Two $U(1)$'s
  and $\epsilon$ Charge Shifts}},}\ }\href {\doibase
  10.1016/0370-2693(86)91377-8} {\bibfield  {journal} {\bibinfo  {journal}
  {Phys. Lett. B}\ }\textbf {\bibinfo {volume} {166}},\ \bibinfo {pages}
  {196--198} (\bibinfo {year} {1986}{\natexlab{b}})}\BibitemShut {NoStop}%
\bibitem [{\citenamefont {Sikivie}(1983)}]{Sikivie:1983ip}%
  \BibitemOpen
  \bibfield  {author} {\bibinfo {author} {\bibfnamefont {P.}~\bibnamefont
  {Sikivie}},\ }\bibfield  {title} {\enquote {\bibinfo {title} {{Experimental
  Tests of the Invisible Axion}},}\ }\href {\doibase
  10.1103/PhysRevLett.51.1415} {\bibfield  {journal} {\bibinfo  {journal}
  {Phys. Rev. Lett.}\ }\textbf {\bibinfo {volume} {51}},\ \bibinfo {pages}
  {1415--1417} (\bibinfo {year} {1983})},\ \bibinfo {note} {[Erratum:
  Phys.Rev.Lett. 52, 695 (1984)]}\BibitemShut {NoStop}%
\bibitem [{\citenamefont {Sikivie}(1985)}]{Sikivie:1985yu}%
  \BibitemOpen
  \bibfield  {author} {\bibinfo {author} {\bibfnamefont {Pierre}\ \bibnamefont
  {Sikivie}},\ }\bibfield  {title} {\enquote {\bibinfo {title} {{Detection
  Rates for 'Invisible' Axion Searches}},}\ }\href {\doibase
  10.1103/PhysRevD.36.974} {\bibfield  {journal} {\bibinfo  {journal} {Phys.
  Rev. D}\ }\textbf {\bibinfo {volume} {32}},\ \bibinfo {pages} {2988}
  (\bibinfo {year} {1985})},\ \bibinfo {note} {[Erratum: Phys.Rev.D 36, 974
  (1987)]}\BibitemShut {NoStop}%
\bibitem [{\citenamefont {Wagner}\ \emph {et~al.}(2010)\citenamefont {Wagner},
  \citenamefont {Rybka}, \citenamefont {Hotz}, \citenamefont {Rosenberg},
  \citenamefont {Asztalos}, \citenamefont {Carosi}, \citenamefont {Hagmann},
  \citenamefont {Kinion}, \citenamefont {van Bibber}, \citenamefont {Hoskins}
  \emph {et~al.}}]{Wagner:2010mi}%
  \BibitemOpen
  \bibfield  {author} {\bibinfo {author} {\bibfnamefont {A.}~\bibnamefont
  {Wagner}}, \bibinfo {author} {\bibfnamefont {G.}~\bibnamefont {Rybka}},
  \bibinfo {author} {\bibfnamefont {M.}~\bibnamefont {Hotz}}, \bibinfo {author}
  {\bibfnamefont {L.~J}\ \bibnamefont {Rosenberg}}, \bibinfo {author}
  {\bibfnamefont {S.~J.}\ \bibnamefont {Asztalos}}, \bibinfo {author}
  {\bibfnamefont {G.}~\bibnamefont {Carosi}}, \bibinfo {author} {\bibfnamefont
  {C.}~\bibnamefont {Hagmann}}, \bibinfo {author} {\bibfnamefont
  {D.}~\bibnamefont {Kinion}}, \bibinfo {author} {\bibfnamefont
  {K.}~\bibnamefont {van Bibber}}, \bibinfo {author} {\bibfnamefont
  {J.}~\bibnamefont {Hoskins}},  \emph {et~al.} (\bibinfo {collaboration} {ADMX
  Collaboration}),\ }\bibfield  {title} {\enquote {\bibinfo {title} {{A Search
  for Hidden Sector Photons with ADMX}},}\ }\href {\doibase
  10.1103/PhysRevLett.105.171801} {\bibfield  {journal} {\bibinfo  {journal}
  {Phys. Rev. Lett.}\ }\textbf {\bibinfo {volume} {105}},\ \bibinfo {pages}
  {171801} (\bibinfo {year} {2010})},\ \Eprint {http://arxiv.org/abs/1007.3766}
  {arXiv:1007.3766 [hep-ex]} \BibitemShut {NoStop}%
\bibitem [{\citenamefont {Jewell}\ \emph {et~al.}(2023)\citenamefont {Jewell},
  \citenamefont {Leder}, \citenamefont {Backes}, \citenamefont {Bai},
  \citenamefont {van Bibber}, \citenamefont {Brubaker}, \citenamefont {Cahn},
  \citenamefont {Droster}, \citenamefont {Esmat}, \citenamefont {Ghosh},
  \citenamefont {Graham}, \citenamefont {Hilton}, \citenamefont {Jackson},
  \citenamefont {Laffan}, \citenamefont {Lamoreaux}, \citenamefont {Lehnert},
  \citenamefont {Lewis}, \citenamefont {Malnou}, \citenamefont {Maruyama},
  \citenamefont {Palken}, \citenamefont {Rapidis}, \citenamefont {Ruddy},
  \citenamefont {Simanovskaia}, \citenamefont {Singh}, \citenamefont {Speller},
  \citenamefont {Vale}, \citenamefont {Wang},\ and\ \citenamefont
  {Zhu}}]{haystaccollaboration2023new}%
  \BibitemOpen
  \bibfield  {author} {\bibinfo {author} {\bibfnamefont {M.~J.}\ \bibnamefont
  {Jewell}}, \bibinfo {author} {\bibfnamefont {A.~F.}\ \bibnamefont {Leder}},
  \bibinfo {author} {\bibfnamefont {K.~M.}\ \bibnamefont {Backes}}, \bibinfo
  {author} {\bibfnamefont {Xiran}\ \bibnamefont {Bai}}, \bibinfo {author}
  {\bibfnamefont {K.}~\bibnamefont {van Bibber}}, \bibinfo {author}
  {\bibfnamefont {B.~M.}\ \bibnamefont {Brubaker}}, \bibinfo {author}
  {\bibfnamefont {S.~B.}\ \bibnamefont {Cahn}}, \bibinfo {author}
  {\bibfnamefont {A.}~\bibnamefont {Droster}}, \bibinfo {author} {\bibfnamefont
  {Maryam~H.}\ \bibnamefont {Esmat}}, \bibinfo {author} {\bibfnamefont
  {Sumita}\ \bibnamefont {Ghosh}}, \bibinfo {author} {\bibfnamefont {Eleanor}\
  \bibnamefont {Graham}}, \bibinfo {author} {\bibfnamefont {Gene~C.}\
  \bibnamefont {Hilton}}, \bibinfo {author} {\bibfnamefont {H.}~\bibnamefont
  {Jackson}}, \bibinfo {author} {\bibfnamefont {Claire}\ \bibnamefont
  {Laffan}}, \bibinfo {author} {\bibfnamefont {S.~K.}\ \bibnamefont
  {Lamoreaux}}, \bibinfo {author} {\bibfnamefont {K.~W.}\ \bibnamefont
  {Lehnert}}, \bibinfo {author} {\bibfnamefont {S.~M.}\ \bibnamefont {Lewis}},
  \bibinfo {author} {\bibfnamefont {M.}~\bibnamefont {Malnou}}, \bibinfo
  {author} {\bibfnamefont {R.~H.}\ \bibnamefont {Maruyama}}, \bibinfo {author}
  {\bibfnamefont {D.~A.}\ \bibnamefont {Palken}}, \bibinfo {author}
  {\bibfnamefont {N.~M.}\ \bibnamefont {Rapidis}}, \bibinfo {author}
  {\bibfnamefont {E.~P.}\ \bibnamefont {Ruddy}}, \bibinfo {author}
  {\bibfnamefont {M.}~\bibnamefont {Simanovskaia}}, \bibinfo {author}
  {\bibfnamefont {Sukhman}\ \bibnamefont {Singh}}, \bibinfo {author}
  {\bibfnamefont {D.~H.}\ \bibnamefont {Speller}}, \bibinfo {author}
  {\bibfnamefont {Leila~R.}\ \bibnamefont {Vale}}, \bibinfo {author}
  {\bibfnamefont {H.}~\bibnamefont {Wang}}, \ and\ \bibinfo {author}
  {\bibfnamefont {Yuqi}\ \bibnamefont {Zhu}} (\bibinfo {collaboration}
  {{HAYSTAC} Collaboration}),\ }\bibfield  {title} {\enquote {\bibinfo {title}
  {New results from {HAYSTAC}'s phase {II} operation with a squeezed state
  receiver},}\ }\href {\doibase 10.1103/PhysRevD.107.072007} {\bibfield
  {journal} {\bibinfo  {journal} {Phys. Rev. D}\ }\textbf {\bibinfo {volume}
  {107}},\ \bibinfo {pages} {072007} (\bibinfo {year} {2023})},\ \Eprint
  {http://arxiv.org/abs/2301.09721} {2301.09721} \BibitemShut {NoStop}%
\bibitem [{\citenamefont {Sikivie}\ \emph {et~al.}(2014)\citenamefont
  {Sikivie}, \citenamefont {Sullivan},\ and\ \citenamefont
  {Tanner}}]{Sikivie:2013laa}%
  \BibitemOpen
  \bibfield  {author} {\bibinfo {author} {\bibfnamefont {P.}~\bibnamefont
  {Sikivie}}, \bibinfo {author} {\bibfnamefont {N.}~\bibnamefont {Sullivan}}, \
  and\ \bibinfo {author} {\bibfnamefont {D.~B.}\ \bibnamefont {Tanner}},\
  }\bibfield  {title} {\enquote {\bibinfo {title} {{Proposal for Axion Dark
  Matter Detection Using an LC Circuit}},}\ }\href {\doibase
  10.1103/PhysRevLett.112.131301} {\bibfield  {journal} {\bibinfo  {journal}
  {Phys. Rev. Lett.}\ }\textbf {\bibinfo {volume} {112}},\ \bibinfo {pages}
  {131301} (\bibinfo {year} {2014})},\ \Eprint {http://arxiv.org/abs/1310.8545}
  {arXiv:1310.8545 [hep-ph]} \BibitemShut {NoStop}%
\bibitem [{\citenamefont {Chaudhuri}\ \emph {et~al.}(2015)\citenamefont
  {Chaudhuri}, \citenamefont {Graham}, \citenamefont {Irwin}, \citenamefont
  {Mardon}, \citenamefont {Rajendran},\ and\ \citenamefont
  {Zhao}}]{Chaudhuri:2014dla}%
  \BibitemOpen
  \bibfield  {author} {\bibinfo {author} {\bibfnamefont {Saptarshi}\
  \bibnamefont {Chaudhuri}}, \bibinfo {author} {\bibfnamefont {Peter~W.}\
  \bibnamefont {Graham}}, \bibinfo {author} {\bibfnamefont {Kent}\ \bibnamefont
  {Irwin}}, \bibinfo {author} {\bibfnamefont {Jeremy}\ \bibnamefont {Mardon}},
  \bibinfo {author} {\bibfnamefont {Surjeet}\ \bibnamefont {Rajendran}}, \ and\
  \bibinfo {author} {\bibfnamefont {Yue}\ \bibnamefont {Zhao}},\ }\bibfield
  {title} {\enquote {\bibinfo {title} {{Radio for hidden-photon dark matter
  detection}},}\ }\href {\doibase 10.1103/PhysRevD.92.075012} {\bibfield
  {journal} {\bibinfo  {journal} {Phys. Rev. D}\ }\textbf {\bibinfo {volume}
  {92}},\ \bibinfo {pages} {075012} (\bibinfo {year} {2015})},\ \Eprint
  {http://arxiv.org/abs/1411.7382} {arXiv:1411.7382 [hep-ph]} \BibitemShut
  {NoStop}%
\bibitem [{\citenamefont {Kahn}\ \emph {et~al.}(2016)\citenamefont {Kahn},
  \citenamefont {Safdi},\ and\ \citenamefont {Thaler}}]{Kahn:2016aff}%
  \BibitemOpen
  \bibfield  {author} {\bibinfo {author} {\bibfnamefont {Yonatan}\ \bibnamefont
  {Kahn}}, \bibinfo {author} {\bibfnamefont {Benjamin~R.}\ \bibnamefont
  {Safdi}}, \ and\ \bibinfo {author} {\bibfnamefont {Jesse}\ \bibnamefont
  {Thaler}},\ }\bibfield  {title} {\enquote {\bibinfo {title} {{Broadband and
  Resonant Approaches to Axion Dark Matter Detection}},}\ }\href {\doibase
  10.1103/PhysRevLett.117.141801} {\bibfield  {journal} {\bibinfo  {journal}
  {Phys. Rev. Lett.}\ }\textbf {\bibinfo {volume} {117}},\ \bibinfo {pages}
  {141801} (\bibinfo {year} {2016})},\ \Eprint
  {http://arxiv.org/abs/1602.01086} {arXiv:1602.01086 [hep-ph]} \BibitemShut
  {NoStop}%
\bibitem [{\citenamefont {Salemi}\ \emph {et~al.}(2021)\citenamefont {Salemi}
  \emph {et~al.}}]{Salemi:2021gck}%
  \BibitemOpen
  \bibfield  {author} {\bibinfo {author} {\bibfnamefont {Chiara~P.}\
  \bibnamefont {Salemi}} \emph {et~al.},\ }\bibfield  {title} {\enquote
  {\bibinfo {title} {{Search for Low-Mass Axion Dark Matter with
  ABRACADABRA-10~cm}},}\ }\href {\doibase 10.1103/PhysRevLett.127.081801}
  {\bibfield  {journal} {\bibinfo  {journal} {Phys. Rev. Lett.}\ }\textbf
  {\bibinfo {volume} {127}},\ \bibinfo {pages} {081801} (\bibinfo {year}
  {2021})},\ \Eprint {http://arxiv.org/abs/2102.06722} {arXiv:2102.06722
  [hep-ex]} \BibitemShut {NoStop}%
\bibitem [{\citenamefont {Braginsky}(1967)}]{braginsky1967classical}%
  \BibitemOpen
  \bibfield  {author} {\bibinfo {author} {\bibfnamefont {VB}~\bibnamefont
  {Braginsky}},\ }\bibfield  {title} {\enquote {\bibinfo {title} {Classical and
  quantum restrictions on the detection of weak disturbances of a macroscopic
  oscillator},}\ }\href@noop {} {\bibfield  {journal} {\bibinfo  {journal} {Zh.
  Eksp. Teor. Fiz}\ }\textbf {\bibinfo {volume} {53}},\ \bibinfo {pages}
  {1434--1441} (\bibinfo {year} {1967})}\BibitemShut {NoStop}%
\bibitem [{\citenamefont {Backes}\ \emph {et~al.}(2021)\citenamefont {Backes}
  \emph {et~al.}}]{HAYSTAC:2020kwv}%
  \BibitemOpen
  \bibfield  {author} {\bibinfo {author} {\bibfnamefont {K.~M.}\ \bibnamefont
  {Backes}} \emph {et~al.} (\bibinfo {collaboration} {HAYSTAC}),\ }\bibfield
  {title} {\enquote {\bibinfo {title} {{A quantum-enhanced search for dark
  matter axions}},}\ }\href {\doibase 10.1038/s41586-021-03226-7} {\bibfield
  {journal} {\bibinfo  {journal} {Nature}\ }\textbf {\bibinfo {volume} {590}},\
  \bibinfo {pages} {238--242} (\bibinfo {year} {2021})},\ \Eprint
  {http://arxiv.org/abs/2008.01853} {arXiv:2008.01853 [quant-ph]} \BibitemShut
  {NoStop}%
\bibitem [{\citenamefont {Chen}\ \emph {et~al.}(2022)\citenamefont {Chen},
  \citenamefont {Jiang}, \citenamefont {Ma}, \citenamefont {Shu},\ and\
  \citenamefont {Yang}}]{Chen:2021bgy}%
  \BibitemOpen
  \bibfield  {author} {\bibinfo {author} {\bibfnamefont {Yifan}\ \bibnamefont
  {Chen}}, \bibinfo {author} {\bibfnamefont {Minyuan}\ \bibnamefont {Jiang}},
  \bibinfo {author} {\bibfnamefont {Yiqiu}\ \bibnamefont {Ma}}, \bibinfo
  {author} {\bibfnamefont {Jing}\ \bibnamefont {Shu}}, \ and\ \bibinfo {author}
  {\bibfnamefont {Yuting}\ \bibnamefont {Yang}},\ }\bibfield  {title} {\enquote
  {\bibinfo {title} {{Axion haloscope array with PT symmetry}},}\ }\href
  {\doibase 10.1103/PhysRevResearch.4.023015} {\bibfield  {journal} {\bibinfo
  {journal} {Phys. Rev. Res.}\ }\textbf {\bibinfo {volume} {4}},\ \bibinfo
  {pages} {023015} (\bibinfo {year} {2022})},\ \Eprint
  {http://arxiv.org/abs/2103.12085} {arXiv:2103.12085 [hep-ph]} \BibitemShut
  {NoStop}%
\bibitem [{\citenamefont {Wurtz}\ \emph {et~al.}(2021)\citenamefont {Wurtz},
  \citenamefont {Brubaker}, \citenamefont {Jiang}, \citenamefont {Ruddy},
  \citenamefont {Palken},\ and\ \citenamefont {Lehnert}}]{Wurtz:2021cnm}%
  \BibitemOpen
  \bibfield  {author} {\bibinfo {author} {\bibfnamefont {K.}~\bibnamefont
  {Wurtz}}, \bibinfo {author} {\bibfnamefont {B.~M.}\ \bibnamefont {Brubaker}},
  \bibinfo {author} {\bibfnamefont {Y.}~\bibnamefont {Jiang}}, \bibinfo
  {author} {\bibfnamefont {E.~P.}\ \bibnamefont {Ruddy}}, \bibinfo {author}
  {\bibfnamefont {D.~A.}\ \bibnamefont {Palken}}, \ and\ \bibinfo {author}
  {\bibfnamefont {K.~W.}\ \bibnamefont {Lehnert}},\ }\bibfield  {title}
  {\enquote {\bibinfo {title} {{Cavity Entanglement and State Swapping to
  Accelerate the Search for Axion Dark Matter}},}\ }\href {\doibase
  10.1103/PRXQuantum.2.040350} {\bibfield  {journal} {\bibinfo  {journal} {PRX
  Quantum}\ }\textbf {\bibinfo {volume} {2}},\ \bibinfo {pages} {040350}
  (\bibinfo {year} {2021})},\ \Eprint {http://arxiv.org/abs/2107.04147}
  {arXiv:2107.04147 [quant-ph]} \BibitemShut {NoStop}%
\bibitem [{\citenamefont {Chen}\ \emph
  {et~al.}(2023{\natexlab{a}})\citenamefont {Chen}, \citenamefont {Li},
  \citenamefont {Liu}, \citenamefont {Shu}, \citenamefont {Yang},\ and\
  \citenamefont {Zeng}}]{Chen:2023ryb}%
  \BibitemOpen
  \bibfield  {author} {\bibinfo {author} {\bibfnamefont {Yifan}\ \bibnamefont
  {Chen}}, \bibinfo {author} {\bibfnamefont {Chunlong}\ \bibnamefont {Li}},
  \bibinfo {author} {\bibfnamefont {Yuxin}\ \bibnamefont {Liu}}, \bibinfo
  {author} {\bibfnamefont {Jing}\ \bibnamefont {Shu}}, \bibinfo {author}
  {\bibfnamefont {Yuting}\ \bibnamefont {Yang}}, \ and\ \bibinfo {author}
  {\bibfnamefont {Yanjie}\ \bibnamefont {Zeng}},\ }\bibfield  {title} {\enquote
  {\bibinfo {title} {{Simultaneous Resonant and Broadband Detection for Dark
  Sectors}},}\ }\href@noop {} {\  (\bibinfo {year} {2023}{\natexlab{a}})},\
  \Eprint {http://arxiv.org/abs/2309.12387} {arXiv:2309.12387 [hep-ph]}
  \BibitemShut {NoStop}%
\bibitem [{\citenamefont {Jiang}\ \emph {et~al.}(2023)\citenamefont {Jiang},
  \citenamefont {Ruddy}, \citenamefont {Quinlan}, \citenamefont {Malnou},
  \citenamefont {Frattini},\ and\ \citenamefont {Lehnert}}]{Jiang:2022vpm}%
  \BibitemOpen
  \bibfield  {author} {\bibinfo {author} {\bibfnamefont {Yue}\ \bibnamefont
  {Jiang}}, \bibinfo {author} {\bibfnamefont {Elizabeth~P.}\ \bibnamefont
  {Ruddy}}, \bibinfo {author} {\bibfnamefont {Kyle~O.}\ \bibnamefont
  {Quinlan}}, \bibinfo {author} {\bibfnamefont {Maxime}\ \bibnamefont
  {Malnou}}, \bibinfo {author} {\bibfnamefont {Nicholas~E.}\ \bibnamefont
  {Frattini}}, \ and\ \bibinfo {author} {\bibfnamefont {Konrad~W.}\
  \bibnamefont {Lehnert}},\ }\bibfield  {title} {\enquote {\bibinfo {title}
  {{Accelerated Weak Signal Search Using Mode Entanglement and State
  Swapping}},}\ }\href {\doibase 10.1103/PRXQuantum.4.020302} {\bibfield
  {journal} {\bibinfo  {journal} {PRX Quantum}\ }\textbf {\bibinfo {volume}
  {4}},\ \bibinfo {pages} {020302} (\bibinfo {year} {2023})},\ \Eprint
  {http://arxiv.org/abs/2211.10403} {arXiv:2211.10403 [quant-ph]} \BibitemShut
  {NoStop}%
\bibitem [{\citenamefont {Dixit}\ \emph {et~al.}(2021)\citenamefont {Dixit},
  \citenamefont {Chakram}, \citenamefont {He}, \citenamefont {Agrawal},
  \citenamefont {Naik}, \citenamefont {Schuster},\ and\ \citenamefont
  {Chou}}]{Dixit:2020ymh}%
  \BibitemOpen
  \bibfield  {author} {\bibinfo {author} {\bibfnamefont {Akash~V.}\
  \bibnamefont {Dixit}}, \bibinfo {author} {\bibfnamefont {Srivatsan}\
  \bibnamefont {Chakram}}, \bibinfo {author} {\bibfnamefont {Kevin}\
  \bibnamefont {He}}, \bibinfo {author} {\bibfnamefont {Ankur}\ \bibnamefont
  {Agrawal}}, \bibinfo {author} {\bibfnamefont {Ravi~K.}\ \bibnamefont {Naik}},
  \bibinfo {author} {\bibfnamefont {David~I.}\ \bibnamefont {Schuster}}, \ and\
  \bibinfo {author} {\bibfnamefont {Aaron}\ \bibnamefont {Chou}},\ }\bibfield
  {title} {\enquote {\bibinfo {title} {{Searching for Dark Matter with a
  Superconducting Qubit}},}\ }\href {\doibase 10.1103/PhysRevLett.126.141302}
  {\bibfield  {journal} {\bibinfo  {journal} {Phys. Rev. Lett.}\ }\textbf
  {\bibinfo {volume} {126}},\ \bibinfo {pages} {141302} (\bibinfo {year}
  {2021})},\ \Eprint {http://arxiv.org/abs/2008.12231} {arXiv:2008.12231
  [hep-ex]} \BibitemShut {NoStop}%
\bibitem [{\citenamefont {Agrawal}\ \emph {et~al.}(2024)\citenamefont
  {Agrawal}, \citenamefont {Dixit}, \citenamefont {Roy}, \citenamefont
  {Chakram}, \citenamefont {He}, \citenamefont {Naik}, \citenamefont
  {Schuster},\ and\ \citenamefont {Chou}}]{Agrawal:2023umy}%
  \BibitemOpen
  \bibfield  {author} {\bibinfo {author} {\bibfnamefont {Ankur}\ \bibnamefont
  {Agrawal}}, \bibinfo {author} {\bibfnamefont {Akash~V}\ \bibnamefont
  {Dixit}}, \bibinfo {author} {\bibfnamefont {Tanay}\ \bibnamefont {Roy}},
  \bibinfo {author} {\bibfnamefont {Srivatsan}\ \bibnamefont {Chakram}},
  \bibinfo {author} {\bibfnamefont {Kevin}\ \bibnamefont {He}}, \bibinfo
  {author} {\bibfnamefont {Ravi~K}\ \bibnamefont {Naik}}, \bibinfo {author}
  {\bibfnamefont {David~I}\ \bibnamefont {Schuster}}, \ and\ \bibinfo {author}
  {\bibfnamefont {Aaron}\ \bibnamefont {Chou}},\ }\bibfield  {title} {\enquote
  {\bibinfo {title} {Stimulated {{Emission}} of {{Signal Photons}} from {{Dark
  Matter Waves}}},}\ }\href {\doibase 10.1103/PhysRevLett.132.140801}
  {\bibfield  {journal} {\bibinfo  {journal} {Phys. Rev. Lett.}\ } (\bibinfo
  {year} {2024}),\ 10.1103/PhysRevLett.132.140801}\BibitemShut {NoStop}%
\bibitem [{\citenamefont {Caves}\ \emph {et~al.}(1980)\citenamefont {Caves},
  \citenamefont {Thorne}, \citenamefont {Drever}, \citenamefont {Sandberg},\
  and\ \citenamefont {Zimmermann}}]{RevModPhys.52.341}%
  \BibitemOpen
  \bibfield  {author} {\bibinfo {author} {\bibfnamefont {Carlton~M.}\
  \bibnamefont {Caves}}, \bibinfo {author} {\bibfnamefont {Kip~S.}\
  \bibnamefont {Thorne}}, \bibinfo {author} {\bibfnamefont {Ronald W.~P.}\
  \bibnamefont {Drever}}, \bibinfo {author} {\bibfnamefont {Vernon~D.}\
  \bibnamefont {Sandberg}}, \ and\ \bibinfo {author} {\bibfnamefont {Mark}\
  \bibnamefont {Zimmermann}},\ }\bibfield  {title} {\enquote {\bibinfo {title}
  {On the measurement of a weak classical force coupled to a quantum-mechanical
  oscillator. i. issues of principle},}\ }\href {\doibase
  10.1103/RevModPhys.52.341} {\bibfield  {journal} {\bibinfo  {journal} {Rev.
  Mod. Phys.}\ }\textbf {\bibinfo {volume} {52}},\ \bibinfo {pages} {341--392}
  (\bibinfo {year} {1980})}\BibitemShut {NoStop}%
\bibitem [{\citenamefont {Braginsky}\ and\ \citenamefont
  {Khalili}(1996)}]{RevModPhys.68.1}%
  \BibitemOpen
  \bibfield  {author} {\bibinfo {author} {\bibfnamefont {V.~B.}\ \bibnamefont
  {Braginsky}}\ and\ \bibinfo {author} {\bibfnamefont {F.~Ya.}\ \bibnamefont
  {Khalili}},\ }\bibfield  {title} {\enquote {\bibinfo {title} {Quantum
  nondemolition measurements: the route from toys to tools},}\ }\href {\doibase
  10.1103/RevModPhys.68.1} {\bibfield  {journal} {\bibinfo  {journal} {Rev.
  Mod. Phys.}\ }\textbf {\bibinfo {volume} {68}},\ \bibinfo {pages} {1--11}
  (\bibinfo {year} {1996})}\BibitemShut {NoStop}%
\bibitem [{\citenamefont {Brady}\ \emph {et~al.}(2022)\citenamefont {Brady},
  \citenamefont {Gao}, \citenamefont {Harnik}, \citenamefont {Liu},
  \citenamefont {Zhang},\ and\ \citenamefont {Zhuang}}]{Brady:2022bus}%
  \BibitemOpen
  \bibfield  {author} {\bibinfo {author} {\bibfnamefont {Anthony~J.}\
  \bibnamefont {Brady}}, \bibinfo {author} {\bibfnamefont {Christina}\
  \bibnamefont {Gao}}, \bibinfo {author} {\bibfnamefont {Roni}\ \bibnamefont
  {Harnik}}, \bibinfo {author} {\bibfnamefont {Zhen}\ \bibnamefont {Liu}},
  \bibinfo {author} {\bibfnamefont {Zheshen}\ \bibnamefont {Zhang}}, \ and\
  \bibinfo {author} {\bibfnamefont {Quntao}\ \bibnamefont {Zhuang}},\
  }\bibfield  {title} {\enquote {\bibinfo {title} {{Entangled Sensor-Networks
  for Dark-Matter Searches}},}\ }\href {\doibase 10.1103/PRXQuantum.3.030333}
  {\bibfield  {journal} {\bibinfo  {journal} {PRX Quantum}\ }\textbf {\bibinfo
  {volume} {3}},\ \bibinfo {pages} {030333} (\bibinfo {year} {2022})},\ \Eprint
  {http://arxiv.org/abs/2203.05375} {arXiv:2203.05375 [quant-ph]} \BibitemShut
  {NoStop}%
\bibitem [{\citenamefont {Brady}\ \emph {et~al.}(2023)\citenamefont {Brady}
  \emph {et~al.}}]{Brady:2022qne}%
  \BibitemOpen
  \bibfield  {author} {\bibinfo {author} {\bibfnamefont {Anthony~J.}\
  \bibnamefont {Brady}} \emph {et~al.},\ }\bibfield  {title} {\enquote
  {\bibinfo {title} {{Entanglement-enhanced optomechanical sensor array with
  application to dark matter searches}},}\ }\href {\doibase
  10.1038/s42005-023-01357-z} {\bibfield  {journal} {\bibinfo  {journal}
  {Commun. Phys.}\ }\textbf {\bibinfo {volume} {6}},\ \bibinfo {pages} {237}
  (\bibinfo {year} {2023})},\ \Eprint {http://arxiv.org/abs/2210.07291}
  {arXiv:2210.07291 [quant-ph]} \BibitemShut {NoStop}%
\bibitem [{\citenamefont {Chen}\ \emph
  {et~al.}(2023{\natexlab{b}})\citenamefont {Chen}, \citenamefont {Fukuda},
  \citenamefont {Inada}, \citenamefont {Moroi}, \citenamefont {Nitta},\ and\
  \citenamefont {Sichanugrist}}]{Chen:2023swh}%
  \BibitemOpen
  \bibfield  {author} {\bibinfo {author} {\bibfnamefont {Shion}\ \bibnamefont
  {Chen}}, \bibinfo {author} {\bibfnamefont {Hajime}\ \bibnamefont {Fukuda}},
  \bibinfo {author} {\bibfnamefont {Toshiaki}\ \bibnamefont {Inada}}, \bibinfo
  {author} {\bibfnamefont {Takeo}\ \bibnamefont {Moroi}}, \bibinfo {author}
  {\bibfnamefont {Tatsumi}\ \bibnamefont {Nitta}}, \ and\ \bibinfo {author}
  {\bibfnamefont {Thanaporn}\ \bibnamefont {Sichanugrist}},\ }\bibfield
  {title} {\enquote {\bibinfo {title} {{Quantum Enhancement in Dark Matter
  Detection with Quantum Computation}},}\ }\href@noop {} {\  (\bibinfo {year}
  {2023}{\natexlab{b}})},\ \Eprint {http://arxiv.org/abs/2311.10413}
  {arXiv:2311.10413 [hep-ph]} \BibitemShut {NoStop}%
\bibitem [{\citenamefont {Ito}\ \emph {et~al.}(2023)\citenamefont {Ito},
  \citenamefont {Kitano}, \citenamefont {Nakano},\ and\ \citenamefont
  {Takai}}]{Ito:2023zhp}%
  \BibitemOpen
  \bibfield  {author} {\bibinfo {author} {\bibfnamefont {Asuka}\ \bibnamefont
  {Ito}}, \bibinfo {author} {\bibfnamefont {Ryuichiro}\ \bibnamefont {Kitano}},
  \bibinfo {author} {\bibfnamefont {Wakutaka}\ \bibnamefont {Nakano}}, \ and\
  \bibinfo {author} {\bibfnamefont {Ryoto}\ \bibnamefont {Takai}},\ }\bibfield
  {title} {\enquote {\bibinfo {title} {{Quantum entanglement of ions for light
  dark matter detection}},}\ }\href@noop {} {\  (\bibinfo {year} {2023})},\
  \Eprint {http://arxiv.org/abs/2311.11632} {arXiv:2311.11632 [hep-ph]}
  \BibitemShut {NoStop}%
\bibitem [{\citenamefont {Ni}\ \emph {et~al.}(2023)\citenamefont {Ni},
  \citenamefont {Li}, \citenamefont {Deng}, \citenamefont {Cai}, \citenamefont
  {Zhang}, \citenamefont {Wang}, \citenamefont {Yang}, \citenamefont {Yu},
  \citenamefont {Yan}, \citenamefont {Liu}, \citenamefont {Zou}, \citenamefont
  {Sun}, \citenamefont {Zheng}, \citenamefont {Xu},\ and\ \citenamefont
  {Yu}}]{ni2023}%
  \BibitemOpen
  \bibfield  {author} {\bibinfo {author} {\bibfnamefont {Zhongchu}\
  \bibnamefont {Ni}}, \bibinfo {author} {\bibfnamefont {Sai}\ \bibnamefont
  {Li}}, \bibinfo {author} {\bibfnamefont {Xiaowei}\ \bibnamefont {Deng}},
  \bibinfo {author} {\bibfnamefont {Yanyan}\ \bibnamefont {Cai}}, \bibinfo
  {author} {\bibfnamefont {Libo}\ \bibnamefont {Zhang}}, \bibinfo {author}
  {\bibfnamefont {Weiting}\ \bibnamefont {Wang}}, \bibinfo {author}
  {\bibfnamefont {Zhen-Biao}\ \bibnamefont {Yang}}, \bibinfo {author}
  {\bibfnamefont {Haifeng}\ \bibnamefont {Yu}}, \bibinfo {author}
  {\bibfnamefont {Fei}\ \bibnamefont {Yan}}, \bibinfo {author} {\bibfnamefont
  {Song}\ \bibnamefont {Liu}}, \bibinfo {author} {\bibfnamefont {Chang-Ling}\
  \bibnamefont {Zou}}, \bibinfo {author} {\bibfnamefont {Luyan}\ \bibnamefont
  {Sun}}, \bibinfo {author} {\bibfnamefont {Shi-Biao}\ \bibnamefont {Zheng}},
  \bibinfo {author} {\bibfnamefont {Yuan}\ \bibnamefont {Xu}}, \ and\ \bibinfo
  {author} {\bibfnamefont {Dapeng}\ \bibnamefont {Yu}},\ }\bibfield  {title}
  {\enquote {\bibinfo {title} {Beating the break-even point with a
  discrete-variable-encoded logical qubit},}\ }\href {\doibase
  10.1038/s41586-023-05784-4} {\bibfield  {journal} {\bibinfo  {journal}
  {Nature}\ }\textbf {\bibinfo {volume} {616}},\ \bibinfo {pages} {56--60}
  (\bibinfo {year} {2023})}\BibitemShut {NoStop}%
\bibitem [{\citenamefont {Lu}\ \emph {et~al.}(2023)\citenamefont {Lu},
  \citenamefont {Maiti}, \citenamefont {Garmon}, \citenamefont {Ganjam},
  \citenamefont {Zhang}, \citenamefont {Claes}, \citenamefont {Frunzio},
  \citenamefont {Girvin},\ and\ \citenamefont {Schoelkopf}}]{lu2023high}%
  \BibitemOpen
  \bibfield  {author} {\bibinfo {author} {\bibfnamefont {Yao}\ \bibnamefont
  {Lu}}, \bibinfo {author} {\bibfnamefont {Aniket}\ \bibnamefont {Maiti}},
  \bibinfo {author} {\bibfnamefont {John~WO}\ \bibnamefont {Garmon}}, \bibinfo
  {author} {\bibfnamefont {Suhas}\ \bibnamefont {Ganjam}}, \bibinfo {author}
  {\bibfnamefont {Yaxing}\ \bibnamefont {Zhang}}, \bibinfo {author}
  {\bibfnamefont {Jahan}\ \bibnamefont {Claes}}, \bibinfo {author}
  {\bibfnamefont {Luigi}\ \bibnamefont {Frunzio}}, \bibinfo {author}
  {\bibfnamefont {Steven~M}\ \bibnamefont {Girvin}}, \ and\ \bibinfo {author}
  {\bibfnamefont {Robert~J}\ \bibnamefont {Schoelkopf}},\ }\bibfield  {title}
  {\enquote {\bibinfo {title} {High-fidelity parametric beamsplitting with a
  parity-protected converter},}\ }\href@noop {} {\bibfield  {journal} {\bibinfo
   {journal} {Nat. Commun.}\ }\textbf {\bibinfo {volume} {14}},\ \bibinfo
  {pages} {5767} (\bibinfo {year} {2023})}\BibitemShut {NoStop}%
\bibitem [{\citenamefont {Jin}\ \emph {et~al.}(2015)\citenamefont {Jin},
  \citenamefont {Kamal}, \citenamefont {Sears}, \citenamefont {Gudmundsen},
  \citenamefont {Hover}, \citenamefont {Miloshi}, \citenamefont {Slattery},
  \citenamefont {Yan}, \citenamefont {Yoder}, \citenamefont {Orlando} \emph
  {et~al.}}]{jin2015thermal}%
  \BibitemOpen
  \bibfield  {author} {\bibinfo {author} {\bibfnamefont {XY}~\bibnamefont
  {Jin}}, \bibinfo {author} {\bibfnamefont {A}~\bibnamefont {Kamal}}, \bibinfo
  {author} {\bibfnamefont {AP}~\bibnamefont {Sears}}, \bibinfo {author}
  {\bibfnamefont {T}~\bibnamefont {Gudmundsen}}, \bibinfo {author}
  {\bibfnamefont {D}~\bibnamefont {Hover}}, \bibinfo {author} {\bibfnamefont
  {J}~\bibnamefont {Miloshi}}, \bibinfo {author} {\bibfnamefont
  {R}~\bibnamefont {Slattery}}, \bibinfo {author} {\bibfnamefont
  {F}~\bibnamefont {Yan}}, \bibinfo {author} {\bibfnamefont {J}~\bibnamefont
  {Yoder}}, \bibinfo {author} {\bibfnamefont {TP}~\bibnamefont {Orlando}},
  \emph {et~al.},\ }\bibfield  {title} {\enquote {\bibinfo {title} {Thermal and
  residual excited-state population in a 3d transmon qubit},}\ }\href@noop {}
  {\bibfield  {journal} {\bibinfo  {journal} {Phys. Rev. Lett.}\ }\textbf
  {\bibinfo {volume} {114}},\ \bibinfo {pages} {240501} (\bibinfo {year}
  {2015})}\BibitemShut {NoStop}%
\bibitem [{\citenamefont {Serniak}\ \emph {et~al.}(2018)\citenamefont
  {Serniak}, \citenamefont {Hays}, \citenamefont {De~Lange}, \citenamefont
  {Diamond}, \citenamefont {Shankar}, \citenamefont {Burkhart}, \citenamefont
  {Frunzio}, \citenamefont {Houzet},\ and\ \citenamefont
  {Devoret}}]{serniak2018hot}%
  \BibitemOpen
  \bibfield  {author} {\bibinfo {author} {\bibfnamefont {K}~\bibnamefont
  {Serniak}}, \bibinfo {author} {\bibfnamefont {M}~\bibnamefont {Hays}},
  \bibinfo {author} {\bibfnamefont {G}~\bibnamefont {De~Lange}}, \bibinfo
  {author} {\bibfnamefont {S}~\bibnamefont {Diamond}}, \bibinfo {author}
  {\bibfnamefont {Sh}~\bibnamefont {Shankar}}, \bibinfo {author} {\bibfnamefont
  {LD}~\bibnamefont {Burkhart}}, \bibinfo {author} {\bibfnamefont
  {L}~\bibnamefont {Frunzio}}, \bibinfo {author} {\bibfnamefont
  {M}~\bibnamefont {Houzet}}, \ and\ \bibinfo {author} {\bibfnamefont
  {MH}~\bibnamefont {Devoret}},\ }\bibfield  {title} {\enquote {\bibinfo
  {title} {Hot nonequilibrium quasiparticles in transmon qubits},}\ }\href@noop
  {} {\bibfield  {journal} {\bibinfo  {journal} {Phys. Rev. Lett.}\ }\textbf
  {\bibinfo {volume} {121}},\ \bibinfo {pages} {157701} (\bibinfo {year}
  {2018})}\BibitemShut {NoStop}%
\bibitem [{\citenamefont {Arute}\ \emph {et~al.}(2019)\citenamefont {Arute},
  \citenamefont {Arya}, \citenamefont {Babbush}, \citenamefont {Bacon},
  \citenamefont {Bardin}, \citenamefont {Barends}, \citenamefont {Biswas},
  \citenamefont {Boixo}, \citenamefont {Brandao}, \citenamefont {Buell} \emph
  {et~al.}}]{arute2019quantum}%
  \BibitemOpen
  \bibfield  {author} {\bibinfo {author} {\bibfnamefont {Frank}\ \bibnamefont
  {Arute}}, \bibinfo {author} {\bibfnamefont {Kunal}\ \bibnamefont {Arya}},
  \bibinfo {author} {\bibfnamefont {Ryan}\ \bibnamefont {Babbush}}, \bibinfo
  {author} {\bibfnamefont {Dave}\ \bibnamefont {Bacon}}, \bibinfo {author}
  {\bibfnamefont {Joseph~C}\ \bibnamefont {Bardin}}, \bibinfo {author}
  {\bibfnamefont {Rami}\ \bibnamefont {Barends}}, \bibinfo {author}
  {\bibfnamefont {Rupak}\ \bibnamefont {Biswas}}, \bibinfo {author}
  {\bibfnamefont {Sergio}\ \bibnamefont {Boixo}}, \bibinfo {author}
  {\bibfnamefont {Fernando~GSL}\ \bibnamefont {Brandao}}, \bibinfo {author}
  {\bibfnamefont {David~A}\ \bibnamefont {Buell}},  \emph {et~al.},\ }\bibfield
   {title} {\enquote {\bibinfo {title} {Quantum supremacy using a programmable
  superconducting processor},}\ }\href@noop {} {\bibfield  {journal} {\bibinfo
  {journal} {Nature}\ }\textbf {\bibinfo {volume} {574}},\ \bibinfo {pages}
  {505--510} (\bibinfo {year} {2019})}\BibitemShut {NoStop}%
\bibitem [{\citenamefont {Sung}\ \emph {et~al.}(2021)\citenamefont {Sung},
  \citenamefont {Ding}, \citenamefont {Braum{\"u}ller}, \citenamefont
  {Veps{\"a}l{\"a}inen}, \citenamefont {Kannan}, \citenamefont {Kjaergaard},
  \citenamefont {Greene}, \citenamefont {Samach}, \citenamefont {McNally},
  \citenamefont {Kim}, \citenamefont {Melville}, \citenamefont {Niedzielski},
  \citenamefont {Schwartz}, \citenamefont {Yoder}, \citenamefont {Orlando},
  \citenamefont {Gustavsson},\ and\ \citenamefont {Oliver}}]{sung2021}%
  \BibitemOpen
  \bibfield  {author} {\bibinfo {author} {\bibfnamefont {Youngkyu}\
  \bibnamefont {Sung}}, \bibinfo {author} {\bibfnamefont {Leon}\ \bibnamefont
  {Ding}}, \bibinfo {author} {\bibfnamefont {Jochen}\ \bibnamefont
  {Braum{\"u}ller}}, \bibinfo {author} {\bibfnamefont {Antti}\ \bibnamefont
  {Veps{\"a}l{\"a}inen}}, \bibinfo {author} {\bibfnamefont {Bharath}\
  \bibnamefont {Kannan}}, \bibinfo {author} {\bibfnamefont {Morten}\
  \bibnamefont {Kjaergaard}}, \bibinfo {author} {\bibfnamefont {Ami}\
  \bibnamefont {Greene}}, \bibinfo {author} {\bibfnamefont {Gabriel~O.}\
  \bibnamefont {Samach}}, \bibinfo {author} {\bibfnamefont {Chris}\
  \bibnamefont {McNally}}, \bibinfo {author} {\bibfnamefont {David}\
  \bibnamefont {Kim}}, \bibinfo {author} {\bibfnamefont {Alexander}\
  \bibnamefont {Melville}}, \bibinfo {author} {\bibfnamefont {Bethany~M.}\
  \bibnamefont {Niedzielski}}, \bibinfo {author} {\bibfnamefont {Mollie~E.}\
  \bibnamefont {Schwartz}}, \bibinfo {author} {\bibfnamefont {Jonilyn~L.}\
  \bibnamefont {Yoder}}, \bibinfo {author} {\bibfnamefont {Terry~P.}\
  \bibnamefont {Orlando}}, \bibinfo {author} {\bibfnamefont {Simon}\
  \bibnamefont {Gustavsson}}, \ and\ \bibinfo {author} {\bibfnamefont
  {William~D.}\ \bibnamefont {Oliver}},\ }\bibfield  {title} {\enquote
  {\bibinfo {title} {Realization of {{High-Fidelity CZ}} and {{ZZ}} -{{Free
  iSWAP Gates}} with a {{Tunable Coupler}}},}\ }\href {\doibase
  10.1103/PhysRevX.11.021058} {\bibfield  {journal} {\bibinfo  {journal} {Phys.
  Rev. X}\ }\textbf {\bibinfo {volume} {11}},\ \bibinfo {pages} {021058}
  (\bibinfo {year} {2021})}\BibitemShut {NoStop}%
\bibitem [{\citenamefont {Brownnutt}\ \emph {et~al.}(2015)\citenamefont
  {Brownnutt}, \citenamefont {Kumph}, \citenamefont {Rabl},\ and\ \citenamefont
  {Blatt}}]{RevModPhys.87.1419}%
  \BibitemOpen
  \bibfield  {author} {\bibinfo {author} {\bibfnamefont {M.}~\bibnamefont
  {Brownnutt}}, \bibinfo {author} {\bibfnamefont {M.}~\bibnamefont {Kumph}},
  \bibinfo {author} {\bibfnamefont {P.}~\bibnamefont {Rabl}}, \ and\ \bibinfo
  {author} {\bibfnamefont {R.}~\bibnamefont {Blatt}},\ }\bibfield  {title}
  {\enquote {\bibinfo {title} {Ion-trap measurements of electric-field noise
  near surfaces},}\ }\href {\doibase 10.1103/RevModPhys.87.1419} {\bibfield
  {journal} {\bibinfo  {journal} {Rev. Mod. Phys.}\ }\textbf {\bibinfo {volume}
  {87}},\ \bibinfo {pages} {1419--1482} (\bibinfo {year} {2015})}\BibitemShut
  {NoStop}%
\bibitem [{\citenamefont {Ballance}\ \emph {et~al.}(2016)\citenamefont
  {Ballance}, \citenamefont {Harty}, \citenamefont {Linke}, \citenamefont
  {Sepiol},\ and\ \citenamefont {Lucas}}]{PhysRevLett.117.060504}%
  \BibitemOpen
  \bibfield  {author} {\bibinfo {author} {\bibfnamefont {C.~J.}\ \bibnamefont
  {Ballance}}, \bibinfo {author} {\bibfnamefont {T.~P.}\ \bibnamefont {Harty}},
  \bibinfo {author} {\bibfnamefont {N.~M.}\ \bibnamefont {Linke}}, \bibinfo
  {author} {\bibfnamefont {M.~A.}\ \bibnamefont {Sepiol}}, \ and\ \bibinfo
  {author} {\bibfnamefont {D.~M.}\ \bibnamefont {Lucas}},\ }\bibfield  {title}
  {\enquote {\bibinfo {title} {High-fidelity quantum logic gates using
  trapped-ion hyperfine qubits},}\ }\href {\doibase
  10.1103/PhysRevLett.117.060504} {\bibfield  {journal} {\bibinfo  {journal}
  {Phys. Rev. Lett.}\ }\textbf {\bibinfo {volume} {117}},\ \bibinfo {pages}
  {060504} (\bibinfo {year} {2016})}\BibitemShut {NoStop}%
\bibitem [{\citenamefont {Gaebler}\ \emph {et~al.}(2016)\citenamefont
  {Gaebler}, \citenamefont {Tan}, \citenamefont {Lin}, \citenamefont {Wan},
  \citenamefont {Bowler}, \citenamefont {Keith}, \citenamefont {Glancy},
  \citenamefont {Coakley}, \citenamefont {Knill}, \citenamefont {Leibfried},\
  and\ \citenamefont {Wineland}}]{PhysRevLett.117.060505}%
  \BibitemOpen
  \bibfield  {author} {\bibinfo {author} {\bibfnamefont {J.~P.}\ \bibnamefont
  {Gaebler}}, \bibinfo {author} {\bibfnamefont {T.~R.}\ \bibnamefont {Tan}},
  \bibinfo {author} {\bibfnamefont {Y.}~\bibnamefont {Lin}}, \bibinfo {author}
  {\bibfnamefont {Y.}~\bibnamefont {Wan}}, \bibinfo {author} {\bibfnamefont
  {R.}~\bibnamefont {Bowler}}, \bibinfo {author} {\bibfnamefont {A.~C.}\
  \bibnamefont {Keith}}, \bibinfo {author} {\bibfnamefont {S.}~\bibnamefont
  {Glancy}}, \bibinfo {author} {\bibfnamefont {K.}~\bibnamefont {Coakley}},
  \bibinfo {author} {\bibfnamefont {E.}~\bibnamefont {Knill}}, \bibinfo
  {author} {\bibfnamefont {D.}~\bibnamefont {Leibfried}}, \ and\ \bibinfo
  {author} {\bibfnamefont {D.~J.}\ \bibnamefont {Wineland}},\ }\bibfield
  {title} {\enquote {\bibinfo {title} {High-fidelity universal gate set for
  ${^{9}\mathrm{Be}}^{+}$ ion qubits},}\ }\href {\doibase
  10.1103/PhysRevLett.117.060505} {\bibfield  {journal} {\bibinfo  {journal}
  {Phys. Rev. Lett.}\ }\textbf {\bibinfo {volume} {117}},\ \bibinfo {pages}
  {060505} (\bibinfo {year} {2016})}\BibitemShut {NoStop}%
\bibitem [{\citenamefont {Javadi-Abhari}\ \emph {et~al.}(2024)\citenamefont
  {Javadi-Abhari}, \citenamefont {Treinish}, \citenamefont {Krsulich},
  \citenamefont {Wood}, \citenamefont {Lishman}, \citenamefont {Gacon},
  \citenamefont {Martiel}, \citenamefont {Nation}, \citenamefont {Bishop},
  \citenamefont {Cross}, \citenamefont {Johnson},\ and\ \citenamefont
  {Gambetta}}]{qiskit2024}%
  \BibitemOpen
  \bibfield  {author} {\bibinfo {author} {\bibfnamefont {Ali}\ \bibnamefont
  {Javadi-Abhari}}, \bibinfo {author} {\bibfnamefont {Matthew}\ \bibnamefont
  {Treinish}}, \bibinfo {author} {\bibfnamefont {Kevin}\ \bibnamefont
  {Krsulich}}, \bibinfo {author} {\bibfnamefont {Christopher~J.}\ \bibnamefont
  {Wood}}, \bibinfo {author} {\bibfnamefont {Jake}\ \bibnamefont {Lishman}},
  \bibinfo {author} {\bibfnamefont {Julien}\ \bibnamefont {Gacon}}, \bibinfo
  {author} {\bibfnamefont {Simon}\ \bibnamefont {Martiel}}, \bibinfo {author}
  {\bibfnamefont {Paul~D.}\ \bibnamefont {Nation}}, \bibinfo {author}
  {\bibfnamefont {Lev~S.}\ \bibnamefont {Bishop}}, \bibinfo {author}
  {\bibfnamefont {Andrew~W.}\ \bibnamefont {Cross}}, \bibinfo {author}
  {\bibfnamefont {Blake~R.}\ \bibnamefont {Johnson}}, \ and\ \bibinfo {author}
  {\bibfnamefont {Jay~M.}\ \bibnamefont {Gambetta}},\ }\href {\doibase
  10.48550/arXiv.2405.08810} {\enquote {\bibinfo {title} {Quantum computing
  with {Q}iskit},}\ } (\bibinfo {year} {2024}),\ \Eprint
  {http://arxiv.org/abs/2405.08810} {arXiv:2405.08810 [quant-ph]} \BibitemShut
  {NoStop}%
\bibitem [{\citenamefont {Deng}\ \emph {et~al.}(2024)\citenamefont {Deng},
  \citenamefont {Li}, \citenamefont {Chen}, \citenamefont {Ni}, \citenamefont
  {Cai}, \citenamefont {Mai}, \citenamefont {Zhang}, \citenamefont {Zheng},
  \citenamefont {Yu}, \citenamefont {Zou}, \citenamefont {Liu}, \citenamefont
  {Yan}, \citenamefont {Xu},\ and\ \citenamefont {Yu}}]{deng2023}%
  \BibitemOpen
  \bibfield  {author} {\bibinfo {author} {\bibfnamefont {Xiaowei}\ \bibnamefont
  {Deng}}, \bibinfo {author} {\bibfnamefont {Sai}\ \bibnamefont {Li}}, \bibinfo
  {author} {\bibfnamefont {Zi-Jie}\ \bibnamefont {Chen}}, \bibinfo {author}
  {\bibfnamefont {Zhongchu}\ \bibnamefont {Ni}}, \bibinfo {author}
  {\bibfnamefont {Yanyan}\ \bibnamefont {Cai}}, \bibinfo {author}
  {\bibfnamefont {Jiasheng}\ \bibnamefont {Mai}}, \bibinfo {author}
  {\bibfnamefont {Libo}\ \bibnamefont {Zhang}}, \bibinfo {author}
  {\bibfnamefont {Pan}\ \bibnamefont {Zheng}}, \bibinfo {author} {\bibfnamefont
  {Haifeng}\ \bibnamefont {Yu}}, \bibinfo {author} {\bibfnamefont {Chang-Ling}\
  \bibnamefont {Zou}}, \bibinfo {author} {\bibfnamefont {Song}\ \bibnamefont
  {Liu}}, \bibinfo {author} {\bibfnamefont {Fei}\ \bibnamefont {Yan}}, \bibinfo
  {author} {\bibfnamefont {Yuan}\ \bibnamefont {Xu}}, \ and\ \bibinfo {author}
  {\bibfnamefont {Dapeng}\ \bibnamefont {Yu}},\ }\bibfield  {title} {\enquote
  {\bibinfo {title} {{Quantum-enhanced metrology with large Fock states}},}\
  }\href {https://www.nature.com/articles/s41567-024-02619-5} {\bibfield
  {journal} {\bibinfo  {journal} {Nat. Phys.}\ } (\bibinfo {year}
  {2024})}\BibitemShut {NoStop}%
\bibitem [{\citenamefont {Caves}(1981)}]{caves1981quantum}%
  \BibitemOpen
  \bibfield  {author} {\bibinfo {author} {\bibfnamefont {Carlton~M}\
  \bibnamefont {Caves}},\ }\bibfield  {title} {\enquote {\bibinfo {title}
  {Quantum-mechanical noise in an interferometer},}\ }\href@noop {} {\bibfield
  {journal} {\bibinfo  {journal} {Phys. Rev. D}\ }\textbf {\bibinfo {volume}
  {23}},\ \bibinfo {pages} {1693} (\bibinfo {year} {1981})}\BibitemShut
  {NoStop}%
\bibitem [{\citenamefont {Lawrie}\ \emph {et~al.}(2019)\citenamefont {Lawrie},
  \citenamefont {Lett}, \citenamefont {Marino},\ and\ \citenamefont
  {Pooser}}]{lawrie2019quantum}%
  \BibitemOpen
  \bibfield  {author} {\bibinfo {author} {\bibfnamefont {Benjamin~J}\
  \bibnamefont {Lawrie}}, \bibinfo {author} {\bibfnamefont {Paul~D}\
  \bibnamefont {Lett}}, \bibinfo {author} {\bibfnamefont {Alberto~M}\
  \bibnamefont {Marino}}, \ and\ \bibinfo {author} {\bibfnamefont {Raphael~C}\
  \bibnamefont {Pooser}},\ }\bibfield  {title} {\enquote {\bibinfo {title}
  {Quantum sensing with squeezed light},}\ }\href@noop {} {\bibfield  {journal}
  {\bibinfo  {journal} {Acs Photonics}\ }\textbf {\bibinfo {volume} {6}},\
  \bibinfo {pages} {1307--1318} (\bibinfo {year} {2019})}\BibitemShut {NoStop}%
\bibitem [{\citenamefont {Eickbusch}\ \emph {et~al.}(2022)\citenamefont
  {Eickbusch}, \citenamefont {Sivak}, \citenamefont {Ding}, \citenamefont
  {Elder}, \citenamefont {Jha}, \citenamefont {Venkatraman}, \citenamefont
  {Royer}, \citenamefont {Girvin}, \citenamefont {Schoelkopf},\ and\
  \citenamefont {Devoret}}]{eickbusch2022fast}%
  \BibitemOpen
  \bibfield  {author} {\bibinfo {author} {\bibfnamefont {Alec}\ \bibnamefont
  {Eickbusch}}, \bibinfo {author} {\bibfnamefont {Volodymyr}\ \bibnamefont
  {Sivak}}, \bibinfo {author} {\bibfnamefont {Andy~Z}\ \bibnamefont {Ding}},
  \bibinfo {author} {\bibfnamefont {Salvatore~S}\ \bibnamefont {Elder}},
  \bibinfo {author} {\bibfnamefont {Shantanu~R}\ \bibnamefont {Jha}}, \bibinfo
  {author} {\bibfnamefont {Jayameenakshi}\ \bibnamefont {Venkatraman}},
  \bibinfo {author} {\bibfnamefont {Baptiste}\ \bibnamefont {Royer}}, \bibinfo
  {author} {\bibfnamefont {Steven~M}\ \bibnamefont {Girvin}}, \bibinfo {author}
  {\bibfnamefont {Robert~J}\ \bibnamefont {Schoelkopf}}, \ and\ \bibinfo
  {author} {\bibfnamefont {Michel~H}\ \bibnamefont {Devoret}},\ }\bibfield
  {title} {\enquote {\bibinfo {title} {Fast universal control of an oscillator
  with weak dispersive coupling to a qubit},}\ }\href@noop {} {\bibfield
  {journal} {\bibinfo  {journal} {Nat. Phys.}\ }\textbf {\bibinfo {volume}
  {18}},\ \bibinfo {pages} {1464--1469} (\bibinfo {year} {2022})}\BibitemShut
  {NoStop}%
\bibitem [{\citenamefont {Ourjoumtsev}\ \emph {et~al.}(2007)\citenamefont
  {Ourjoumtsev}, \citenamefont {Jeong}, \citenamefont {{Tualle-Brouri}},\ and\
  \citenamefont {Grangier}}]{ourjoumtsev2007}%
  \BibitemOpen
  \bibfield  {author} {\bibinfo {author} {\bibfnamefont {Alexei}\ \bibnamefont
  {Ourjoumtsev}}, \bibinfo {author} {\bibfnamefont {Hyunseok}\ \bibnamefont
  {Jeong}}, \bibinfo {author} {\bibfnamefont {Rosa}\ \bibnamefont
  {{Tualle-Brouri}}}, \ and\ \bibinfo {author} {\bibfnamefont {Philippe}\
  \bibnamefont {Grangier}},\ }\bibfield  {title} {\enquote {\bibinfo {title}
  {Generation of optical `{{Schr{\"o}dinger}} cats' from photon number
  states},}\ }\href {\doibase 10.1038/nature06054} {\bibfield  {journal}
  {\bibinfo  {journal} {Nature}\ }\textbf {\bibinfo {volume} {448}},\ \bibinfo
  {pages} {784--786} (\bibinfo {year} {2007})}\BibitemShut {NoStop}%
\bibitem [{\citenamefont {Vlastakis}\ \emph {et~al.}(2013)\citenamefont
  {Vlastakis}, \citenamefont {Kirchmair}, \citenamefont {Leghtas},
  \citenamefont {Nigg}, \citenamefont {Frunzio}, \citenamefont {Girvin},
  \citenamefont {Mirrahimi}, \citenamefont {Devoret},\ and\ \citenamefont
  {Schoelkopf}}]{vlastakis2013}%
  \BibitemOpen
  \bibfield  {author} {\bibinfo {author} {\bibfnamefont {Brian}\ \bibnamefont
  {Vlastakis}}, \bibinfo {author} {\bibfnamefont {Gerhard}\ \bibnamefont
  {Kirchmair}}, \bibinfo {author} {\bibfnamefont {Zaki}\ \bibnamefont
  {Leghtas}}, \bibinfo {author} {\bibfnamefont {Simon~E.}\ \bibnamefont
  {Nigg}}, \bibinfo {author} {\bibfnamefont {Luigi}\ \bibnamefont {Frunzio}},
  \bibinfo {author} {\bibfnamefont {S.~M.}\ \bibnamefont {Girvin}}, \bibinfo
  {author} {\bibfnamefont {Mazyar}\ \bibnamefont {Mirrahimi}}, \bibinfo
  {author} {\bibfnamefont {M.~H.}\ \bibnamefont {Devoret}}, \ and\ \bibinfo
  {author} {\bibfnamefont {R.~J.}\ \bibnamefont {Schoelkopf}},\ }\bibfield
  {title} {\enquote {\bibinfo {title} {Deterministically {{Encoding Quantum
  Information Using}} 100-{{Photon Schr{\"o}dinger Cat States}}},}\ }\href
  {\doibase 10.1126/science.1243289} {\bibfield  {journal} {\bibinfo  {journal}
  {Science}\ }\textbf {\bibinfo {volume} {342}},\ \bibinfo {pages} {607--610}
  (\bibinfo {year} {2013})}\BibitemShut {NoStop}%
\bibitem [{\citenamefont {Ofek}\ \emph {et~al.}(2016)\citenamefont {Ofek},
  \citenamefont {Petrenko}, \citenamefont {Heeres}, \citenamefont {Reinhold},
  \citenamefont {Leghtas}, \citenamefont {Vlastakis}, \citenamefont {Liu},
  \citenamefont {Frunzio}, \citenamefont {Girvin}, \citenamefont {Jiang} \emph
  {et~al.}}]{ofek2016extending}%
  \BibitemOpen
  \bibfield  {author} {\bibinfo {author} {\bibfnamefont {Nissim}\ \bibnamefont
  {Ofek}}, \bibinfo {author} {\bibfnamefont {Andrei}\ \bibnamefont {Petrenko}},
  \bibinfo {author} {\bibfnamefont {Reinier}\ \bibnamefont {Heeres}}, \bibinfo
  {author} {\bibfnamefont {Philip}\ \bibnamefont {Reinhold}}, \bibinfo {author}
  {\bibfnamefont {Zaki}\ \bibnamefont {Leghtas}}, \bibinfo {author}
  {\bibfnamefont {Brian}\ \bibnamefont {Vlastakis}}, \bibinfo {author}
  {\bibfnamefont {Yehan}\ \bibnamefont {Liu}}, \bibinfo {author} {\bibfnamefont
  {Luigi}\ \bibnamefont {Frunzio}}, \bibinfo {author} {\bibfnamefont
  {SM}~\bibnamefont {Girvin}}, \bibinfo {author} {\bibfnamefont {Liang}\
  \bibnamefont {Jiang}},  \emph {et~al.},\ }\bibfield  {title} {\enquote
  {\bibinfo {title} {Extending the lifetime of a quantum bit with error
  correction in superconducting circuits},}\ }\href@noop {} {\bibfield
  {journal} {\bibinfo  {journal} {Nature}\ }\textbf {\bibinfo {volume} {536}},\
  \bibinfo {pages} {441--445} (\bibinfo {year} {2016})}\BibitemShut {NoStop}%
\bibitem [{\citenamefont {Milul}\ \emph {et~al.}(2023)\citenamefont {Milul},
  \citenamefont {Guttel}, \citenamefont {Goldblatt}, \citenamefont {Hazanov},
  \citenamefont {Joshi}, \citenamefont {Chausovsky}, \citenamefont {Kahn},
  \citenamefont {{\c C}ifty{\"u}rek}, \citenamefont {Lafont},\ and\
  \citenamefont {Rosenblum}}]{milul2023}%
  \BibitemOpen
  \bibfield  {author} {\bibinfo {author} {\bibfnamefont {Ofir}\ \bibnamefont
  {Milul}}, \bibinfo {author} {\bibfnamefont {Barkay}\ \bibnamefont {Guttel}},
  \bibinfo {author} {\bibfnamefont {Uri}\ \bibnamefont {Goldblatt}}, \bibinfo
  {author} {\bibfnamefont {Sergey}\ \bibnamefont {Hazanov}}, \bibinfo {author}
  {\bibfnamefont {Lalit~M.}\ \bibnamefont {Joshi}}, \bibinfo {author}
  {\bibfnamefont {Daniel}\ \bibnamefont {Chausovsky}}, \bibinfo {author}
  {\bibfnamefont {Nitzan}\ \bibnamefont {Kahn}}, \bibinfo {author}
  {\bibfnamefont {Engin}\ \bibnamefont {{\c C}ifty{\"u}rek}}, \bibinfo {author}
  {\bibfnamefont {Fabien}\ \bibnamefont {Lafont}}, \ and\ \bibinfo {author}
  {\bibfnamefont {Serge}\ \bibnamefont {Rosenblum}},\ }\bibfield  {title}
  {\enquote {\bibinfo {title} {Superconducting {{Cavity Qubit}} with {{Tens}}
  of {{Milliseconds Single-Photon Coherence Time}}},}\ }\href {\doibase
  10.1103/PRXQuantum.4.030336} {\bibfield  {journal} {\bibinfo  {journal} {PRX
  Quantum}\ }\textbf {\bibinfo {volume} {4}},\ \bibinfo {pages} {030336}
  (\bibinfo {year} {2023})}\BibitemShut {NoStop}%
\bibitem [{\citenamefont {Pan}\ \emph {et~al.}(2024)\citenamefont {Pan},
  \citenamefont {Krisnanda}, \citenamefont {Duina}, \citenamefont {Park},
  \citenamefont {Song}, \citenamefont {Fontaine}, \citenamefont {Copetudo},
  \citenamefont {Filip},\ and\ \citenamefont {Gao}}]{pan2024}%
  \BibitemOpen
  \bibfield  {author} {\bibinfo {author} {\bibfnamefont {Xiaozhou}\
  \bibnamefont {Pan}}, \bibinfo {author} {\bibfnamefont {Tanjung}\ \bibnamefont
  {Krisnanda}}, \bibinfo {author} {\bibfnamefont {Andrea}\ \bibnamefont
  {Duina}}, \bibinfo {author} {\bibfnamefont {Kimin}\ \bibnamefont {Park}},
  \bibinfo {author} {\bibfnamefont {Pengtao}\ \bibnamefont {Song}}, \bibinfo
  {author} {\bibfnamefont {Clara~Yun}\ \bibnamefont {Fontaine}}, \bibinfo
  {author} {\bibfnamefont {Adrian}\ \bibnamefont {Copetudo}}, \bibinfo {author}
  {\bibfnamefont {Radim}\ \bibnamefont {Filip}}, \ and\ \bibinfo {author}
  {\bibfnamefont {Yvonne~Y.}\ \bibnamefont {Gao}},\ }\bibfield  {title}
  {\enquote {\bibinfo {title} {Realisation of versatile and effective quantum
  metrology using a single bosonic mode},}\ }\href
  {https://arxiv.org/abs/2403.14967} {\  (\bibinfo {year} {2024})},\ \Eprint
  {http://arxiv.org/abs/2403.14967} {arXiv:2403.14967 [quant-ph]} \BibitemShut
  {NoStop}%
\bibitem [{\citenamefont {Berlin}\ \emph {et~al.}(2020)\citenamefont {Berlin},
  \citenamefont {D'Agnolo}, \citenamefont {Ellis}, \citenamefont {Nantista},
  \citenamefont {Neilson}, \citenamefont {Schuster}, \citenamefont {Tantawi},
  \citenamefont {Toro},\ and\ \citenamefont {Zhou}}]{Berlin:2019ahk}%
  \BibitemOpen
  \bibfield  {author} {\bibinfo {author} {\bibfnamefont {Asher}\ \bibnamefont
  {Berlin}}, \bibinfo {author} {\bibfnamefont {Raffaele~Tito}\ \bibnamefont
  {D'Agnolo}}, \bibinfo {author} {\bibfnamefont {Sebastian A.~R.}\ \bibnamefont
  {Ellis}}, \bibinfo {author} {\bibfnamefont {Christopher}\ \bibnamefont
  {Nantista}}, \bibinfo {author} {\bibfnamefont {Jeffrey}\ \bibnamefont
  {Neilson}}, \bibinfo {author} {\bibfnamefont {Philip}\ \bibnamefont
  {Schuster}}, \bibinfo {author} {\bibfnamefont {Sami}\ \bibnamefont
  {Tantawi}}, \bibinfo {author} {\bibfnamefont {Natalia}\ \bibnamefont {Toro}},
  \ and\ \bibinfo {author} {\bibfnamefont {Kevin}\ \bibnamefont {Zhou}},\
  }\bibfield  {title} {\enquote {\bibinfo {title} {{Axion Dark Matter Detection
  by Superconducting Resonant Frequency Conversion}},}\ }\href {\doibase
  10.1007/JHEP07(2020)088} {\bibfield  {journal} {\bibinfo  {journal} {JHEP}\
  }\textbf {\bibinfo {volume} {07}},\ \bibinfo {pages} {088} (\bibinfo {year}
  {2020})},\ \Eprint {http://arxiv.org/abs/1912.11048} {arXiv:1912.11048
  [hep-ph]} \BibitemShut {NoStop}%
\bibitem [{\citenamefont {Berlin}\ \emph {et~al.}(2021)\citenamefont {Berlin},
  \citenamefont {D'Agnolo}, \citenamefont {Ellis},\ and\ \citenamefont
  {Zhou}}]{Berlin:2020vrk}%
  \BibitemOpen
  \bibfield  {author} {\bibinfo {author} {\bibfnamefont {Asher}\ \bibnamefont
  {Berlin}}, \bibinfo {author} {\bibfnamefont {Raffaele~Tito}\ \bibnamefont
  {D'Agnolo}}, \bibinfo {author} {\bibfnamefont {Sebastian A.~R.}\ \bibnamefont
  {Ellis}}, \ and\ \bibinfo {author} {\bibfnamefont {Kevin}\ \bibnamefont
  {Zhou}},\ }\bibfield  {title} {\enquote {\bibinfo {title} {{Heterodyne
  broadband detection of axion dark matter}},}\ }\href {\doibase
  10.1103/PhysRevD.104.L111701} {\bibfield  {journal} {\bibinfo  {journal}
  {Phys. Rev. D}\ }\textbf {\bibinfo {volume} {104}},\ \bibinfo {pages}
  {L111701} (\bibinfo {year} {2021})},\ \Eprint
  {http://arxiv.org/abs/2007.15656} {arXiv:2007.15656 [hep-ph]} \BibitemShut
  {NoStop}%
\bibitem [{\citenamefont {Chen}\ \emph {et~al.}(2024)\citenamefont {Chen},
  \citenamefont {Fukuda}, \citenamefont {Inada}, \citenamefont {Moroi},
  \citenamefont {Nitta},\ and\ \citenamefont {Sichanugrist}}]{Chen:2024aya}%
  \BibitemOpen
  \bibfield  {author} {\bibinfo {author} {\bibfnamefont {Shion}\ \bibnamefont
  {Chen}}, \bibinfo {author} {\bibfnamefont {Hajime}\ \bibnamefont {Fukuda}},
  \bibinfo {author} {\bibfnamefont {Toshiaki}\ \bibnamefont {Inada}}, \bibinfo
  {author} {\bibfnamefont {Takeo}\ \bibnamefont {Moroi}}, \bibinfo {author}
  {\bibfnamefont {Tatsumi}\ \bibnamefont {Nitta}}, \ and\ \bibinfo {author}
  {\bibfnamefont {Thanaporn}\ \bibnamefont {Sichanugrist}},\ }\bibfield
  {title} {\enquote {\bibinfo {title} {{Search for QCD axion dark matter with
  transmon qubits and quantum circuit}},}\ }\href@noop {} {\  (\bibinfo {year}
  {2024})},\ \Eprint {http://arxiv.org/abs/2407.19755} {arXiv:2407.19755
  [hep-ph]} \BibitemShut {NoStop}%
\bibitem [{\citenamefont {Krause}\ \emph {et~al.}(2022)\citenamefont {Krause},
  \citenamefont {Dickel}, \citenamefont {Vaal}, \citenamefont {Vielmetter},
  \citenamefont {Feng}, \citenamefont {Bounds}, \citenamefont {Catelani},
  \citenamefont {Fink},\ and\ \citenamefont {Ando}}]{krause2022magnetic}%
  \BibitemOpen
  \bibfield  {author} {\bibinfo {author} {\bibfnamefont {J}~\bibnamefont
  {Krause}}, \bibinfo {author} {\bibfnamefont {C}~\bibnamefont {Dickel}},
  \bibinfo {author} {\bibfnamefont {E}~\bibnamefont {Vaal}}, \bibinfo {author}
  {\bibfnamefont {M}~\bibnamefont {Vielmetter}}, \bibinfo {author}
  {\bibfnamefont {J}~\bibnamefont {Feng}}, \bibinfo {author} {\bibfnamefont
  {R}~\bibnamefont {Bounds}}, \bibinfo {author} {\bibfnamefont {G}~\bibnamefont
  {Catelani}}, \bibinfo {author} {\bibfnamefont {Johannes~M}\ \bibnamefont
  {Fink}}, \ and\ \bibinfo {author} {\bibfnamefont {Yoichi}\ \bibnamefont
  {Ando}},\ }\bibfield  {title} {\enquote {\bibinfo {title} {Magnetic field
  resilience of three-dimensional transmons with thin-film al/alo x/al
  josephson junctions approaching 1 t},}\ }\href@noop {} {\bibfield  {journal}
  {\bibinfo  {journal} {Phys. Rev. Appl.}\ }\textbf {\bibinfo {volume} {17}},\
  \bibinfo {pages} {034032} (\bibinfo {year} {2022})}\BibitemShut {NoStop}%
\bibitem [{\citenamefont {Rokhinson}\ \emph {et~al.}(2012)\citenamefont
  {Rokhinson}, \citenamefont {Liu},\ and\ \citenamefont
  {Furdyna}}]{rokhinson2012fractional}%
  \BibitemOpen
  \bibfield  {author} {\bibinfo {author} {\bibfnamefont {Leonid~P}\
  \bibnamefont {Rokhinson}}, \bibinfo {author} {\bibfnamefont {Xinyu}\
  \bibnamefont {Liu}}, \ and\ \bibinfo {author} {\bibfnamefont {Jacek~K}\
  \bibnamefont {Furdyna}},\ }\bibfield  {title} {\enquote {\bibinfo {title}
  {The fractional ac josephson effect in a semiconductor--superconductor
  nanowire as a signature of majorana particles},}\ }\href@noop {} {\bibfield
  {journal} {\bibinfo  {journal} {Nat. Phys.}\ }\textbf {\bibinfo {volume}
  {8}},\ \bibinfo {pages} {795--799} (\bibinfo {year} {2012})}\BibitemShut
  {NoStop}%
\bibitem [{\citenamefont {Aggarwal}\ \emph {et~al.}(2021)\citenamefont
  {Aggarwal} \emph {et~al.}}]{Aggarwal:2020olq}%
  \BibitemOpen
  \bibfield  {author} {\bibinfo {author} {\bibfnamefont {Nancy}\ \bibnamefont
  {Aggarwal}} \emph {et~al.},\ }\bibfield  {title} {\enquote {\bibinfo {title}
  {{Challenges and opportunities of gravitational-wave searches at MHz to GHz
  frequencies}},}\ }\href {\doibase 10.1007/s41114-021-00032-5} {\bibfield
  {journal} {\bibinfo  {journal} {Living Rev. Rel.}\ }\textbf {\bibinfo
  {volume} {24}},\ \bibinfo {pages} {4} (\bibinfo {year} {2021})},\ \Eprint
  {http://arxiv.org/abs/2011.12414} {arXiv:2011.12414 [gr-qc]} \BibitemShut
  {NoStop}%
\bibitem [{\citenamefont {Berlin}\ \emph {et~al.}(2022)\citenamefont {Berlin},
  \citenamefont {Blas}, \citenamefont {Tito~D'Agnolo}, \citenamefont {Ellis},
  \citenamefont {Harnik}, \citenamefont {Kahn},\ and\ \citenamefont
  {Sch\"utte-Engel}}]{Berlin:2021txa}%
  \BibitemOpen
  \bibfield  {author} {\bibinfo {author} {\bibfnamefont {Asher}\ \bibnamefont
  {Berlin}}, \bibinfo {author} {\bibfnamefont {Diego}\ \bibnamefont {Blas}},
  \bibinfo {author} {\bibfnamefont {Raffaele}\ \bibnamefont {Tito~D'Agnolo}},
  \bibinfo {author} {\bibfnamefont {Sebastian A.~R.}\ \bibnamefont {Ellis}},
  \bibinfo {author} {\bibfnamefont {Roni}\ \bibnamefont {Harnik}}, \bibinfo
  {author} {\bibfnamefont {Yonatan}\ \bibnamefont {Kahn}}, \ and\ \bibinfo
  {author} {\bibfnamefont {Jan}\ \bibnamefont {Sch\"utte-Engel}},\ }\bibfield
  {title} {\enquote {\bibinfo {title} {{Detecting high-frequency gravitational
  waves with microwave cavities}},}\ }\href {\doibase
  10.1103/PhysRevD.105.116011} {\bibfield  {journal} {\bibinfo  {journal}
  {Phys. Rev. D}\ }\textbf {\bibinfo {volume} {105}},\ \bibinfo {pages}
  {116011} (\bibinfo {year} {2022})},\ \Eprint
  {http://arxiv.org/abs/2112.11465} {arXiv:2112.11465 [hep-ph]} \BibitemShut
  {NoStop}%
\bibitem [{\citenamefont {Hill}(2009)}]{hill2009electromagnetic}%
  \BibitemOpen
  \bibfield  {author} {\bibinfo {author} {\bibfnamefont {David~A}\ \bibnamefont
  {Hill}},\ }\href@noop {} {\emph {\bibinfo {title} {Electromagnetic fields in
  cavities: deterministic and statistical theories}}}\ (\bibinfo  {publisher}
  {John Wiley \& Sons},\ \bibinfo {year} {2009})\BibitemShut {NoStop}%
\bibitem [{\citenamefont {Koch}\ \emph {et~al.}(2007)\citenamefont {Koch},
  \citenamefont {Yu}, \citenamefont {Gambetta}, \citenamefont {Houck},
  \citenamefont {Schuster}, \citenamefont {Majer}, \citenamefont {Blais},
  \citenamefont {Devoret}, \citenamefont {Girvin},\ and\ \citenamefont
  {Schoelkopf}}]{koch2007}%
  \BibitemOpen
  \bibfield  {author} {\bibinfo {author} {\bibfnamefont {Jens}\ \bibnamefont
  {Koch}}, \bibinfo {author} {\bibfnamefont {Terri~M.}\ \bibnamefont {Yu}},
  \bibinfo {author} {\bibfnamefont {Jay}\ \bibnamefont {Gambetta}}, \bibinfo
  {author} {\bibfnamefont {A.~A.}\ \bibnamefont {Houck}}, \bibinfo {author}
  {\bibfnamefont {D.~I.}\ \bibnamefont {Schuster}}, \bibinfo {author}
  {\bibfnamefont {J.}~\bibnamefont {Majer}}, \bibinfo {author} {\bibfnamefont
  {Alexandre}\ \bibnamefont {Blais}}, \bibinfo {author} {\bibfnamefont {M.~H.}\
  \bibnamefont {Devoret}}, \bibinfo {author} {\bibfnamefont {S.~M.}\
  \bibnamefont {Girvin}}, \ and\ \bibinfo {author} {\bibfnamefont {R.~J.}\
  \bibnamefont {Schoelkopf}},\ }\bibfield  {title} {\enquote {\bibinfo {title}
  {Charge-insensitive qubit design derived from the {{Cooper}} pair box},}\
  }\href {\doibase 10.1103/PhysRevA.76.042319} {\bibfield  {journal} {\bibinfo
  {journal} {Phys. Rev. A}\ }\textbf {\bibinfo {volume} {76}},\ \bibinfo
  {pages} {042319} (\bibinfo {year} {2007})}\BibitemShut {NoStop}%
\bibitem [{\citenamefont {Chen}\ \emph
  {et~al.}(2023{\natexlab{c}})\citenamefont {Chen}, \citenamefont {Fukuda},
  \citenamefont {Inada}, \citenamefont {Moroi}, \citenamefont {Nitta},\ and\
  \citenamefont {Sichanugrist}}]{Chen:2022quj}%
  \BibitemOpen
  \bibfield  {author} {\bibinfo {author} {\bibfnamefont {Shion}\ \bibnamefont
  {Chen}}, \bibinfo {author} {\bibfnamefont {Hajime}\ \bibnamefont {Fukuda}},
  \bibinfo {author} {\bibfnamefont {Toshiaki}\ \bibnamefont {Inada}}, \bibinfo
  {author} {\bibfnamefont {Takeo}\ \bibnamefont {Moroi}}, \bibinfo {author}
  {\bibfnamefont {Tatsumi}\ \bibnamefont {Nitta}}, \ and\ \bibinfo {author}
  {\bibfnamefont {Thanaporn}\ \bibnamefont {Sichanugrist}},\ }\bibfield
  {title} {\enquote {\bibinfo {title} {{Detecting Hidden Photon Dark Matter
  Using the Direct Excitation of Transmon Qubits}},}\ }\href {\doibase
  10.1103/PhysRevLett.131.211001} {\bibfield  {journal} {\bibinfo  {journal}
  {Phys. Rev. Lett.}\ }\textbf {\bibinfo {volume} {131}},\ \bibinfo {pages}
  {211001} (\bibinfo {year} {2023}{\natexlab{c}})},\ \Eprint
  {http://arxiv.org/abs/2212.03884} {arXiv:2212.03884 [hep-ph]} \BibitemShut
  {NoStop}%
\bibitem [{\citenamefont {Blank}\ \emph {et~al.}(2020)\citenamefont {Blank},
  \citenamefont {Park}, \citenamefont {Rhee},\ and\ \citenamefont
  {Petruccione}}]{Blank:2020zfq}%
  \BibitemOpen
  \bibfield  {author} {\bibinfo {author} {\bibfnamefont {Carsten}\ \bibnamefont
  {Blank}}, \bibinfo {author} {\bibfnamefont {Daniel~K.}\ \bibnamefont {Park}},
  \bibinfo {author} {\bibfnamefont {June-Koo~Kevin}\ \bibnamefont {Rhee}}, \
  and\ \bibinfo {author} {\bibfnamefont {Francesco}\ \bibnamefont
  {Petruccione}},\ }\bibfield  {title} {\enquote {\bibinfo {title} {{Quantum
  classifier with tailored quantum kernel}},}\ }\href {\doibase
  10.1038/s41534-020-0272-6} {\bibfield  {journal} {\bibinfo  {journal} {npj
  Quantum Inf.}\ }\textbf {\bibinfo {volume} {6}},\ \bibinfo {pages} {41}
  (\bibinfo {year} {2020})}\BibitemShut {NoStop}%
\end{thebibliography}%
\widetext
\begin{center}
	\textbf{\large Supplemental Material: Eliminating Incoherent Noise: A Coherent Quantum Approach in Multi-Sensor Dark Matter Detection}
\end{center}
\setcounter{section}{0}
\setcounter{equation}{0}
\setcounter{figure}{0}
\setcounter{table}{0}
\makeatletter
\renewcommand{\theequation}{S\arabic{equation}}
\renewcommand{\thefigure}{S\arabic{figure}}
\renewcommand{\bibnumfmt}[1]{[#1]}
\renewcommand{\citenumfont}[1]{#1}

\section{Detection schemes with resonant oscillators} \label{sec:detect}

In this section, we focus on the electromagnetic resonant (cavity and superconducting circuit) and mechanical resonant (trapped ion) systems and set up a parameterization for the interaction between quantized resonant modes and potential signals, such as an effective current. 
We follow Ref.~\cite{Chen:2023ryb} for a similar derivation of the Hamiltonian of electromagnetic resonant systems.

\subsection{Resonant cavity}\label{sec:rc}
The electromagnetic fields within a cavity of volume $V$ are quantized bound states.
In the Coulomb gauge, the vector potential can be parameterized as 
\begin{equation}
    \vec{A}  = \sum_n \frac{1}{\sqrt{2\omega_r^n}} \hat{a}_n^\dagger \vec{\epsilon}_n(\vec{r}\,)e^{-\mi\omega_r^n\, t}+ h.c..
    \label{eq:Aq}
\end{equation}
where the sum is taken over various modes labeled by $n$, and $\hat{a}_n$ ($\hat{a}_n^\dagger$) it the annihilation (creation) operator for a mode characterized by an eigenfrequency $\omega_r^n$ and a wave function $\vec{\epsilon}_n(\vec{r}\,)$. 
The electric and magnetic fields of the cavity modes are given by
\begin{equation} 
\vec{E} = -\partial \vec{A}/ \partial t,\qquad 
\vec{B} = \vec{\nabla} \times \vec{A}.
\label{eq:EB}
\end{equation}

Considering a perfect conductor as the boundary, each mode is determined by the vacuum Maxwell’s equation $\Box \vec{A} = 0$ inside the receiver cavity, along with the boundary condition $\vec B_\perp = 0$ and $\vec E_\parallel = 0$. Furthermore, the wave functions must satisfy the orthonormality condition:
\be \int_\text{V}\vec{\epsilon}_m^{\ *} \, \vec{\epsilon}_n\,  \D^3 \vec{r}  = \delta_{mn}.\label{eq:norm} \ee
Using Eqs. (\ref{eq:Aq}-\ref{eq:norm}), the free Hamiltonian for cavity modes can be written as:
\be\begin{split}
    H_0 =  \frac{1}{2} \int_\text{V} \left(\vec{E}^2+\vec{B}^2\right)\, \D^3 \vec{r} = \sum_n \omega_r^n \left(\hat{a}_n^\dagger \hat{a}_n+\frac{1}{2}\right).\label{eq:Hcav}
    \end{split}
\ee

A cavity mode is generated resonantly once its frequency falls within the resonant bandwidth and its wave function spatially overlaps with the effective current $J^\mu_{\rm eff}$ originating from bosonic fields. The interaction Hamiltonian concerning the spatial component of the effective current $\vec{J}_{\rm eff}$ is given by a linear coupling with the vector potential $\vec{A}$:

\be\begin{split} H_{\text{int}} =
    \int_\text{V}\,\vec{A}\cdot \vec{J}_{\rm eff}\, \D^3 \vec{r}
    = \sum_n \sqrt{\frac{V}{2 \omega_r^n}}\, G_{n}\, \bar{J}_{\rm eff}\, \hat{a}_n^{\dagger}\, e^{-\mi \omega_r^n\, t} + h.c. .\label{eq:cavalp}
  \end{split}
\ee
where $G_n$ is the geometric overlap factor between $\vec{\epsilon}_n$ and $\vec{J}_{\rm eff}$:
\begin{equation}
    G_{n} \equiv \frac{{\int}_{V} \, \vec{\epsilon}_{n} \cdot \vec{J}_{\rm eff}\,  \D^3 \vec{r}}{\sqrt{\int_{V}|\vec{J}_{\rm eff}|^{2}\,  \D^3 \vec{r}}}.
    \label{eq:etan}
\end{equation}
and $\bar{J}_{\rm eff}$ denotes the average current density within the cavity:
\begin{equation}
   \bar{J}_{\rm eff} \equiv \sqrt{ \frac{1}{V}\int_{V}|\vec{J}_{\rm eff}|^{2} \,  \D^3 \vec{r}}.\label{eq:jeffint}
\end{equation}

On the other hand, the time component of $J^\mu_{\rm eff}$, corresponding to an effective charge density $\rho_{\rm eff}$, exclusively excites the irrotational mode~\cite{hill2009electromagnetic}, which does not experience resonant enhancement within the cavity context.

\subsection{Superconducting Quantum Circuits}
Before we introduce the nonlinear quantum circuits, we first quantize the Hamiltonian of an LC resonator characterized by its inductance $L$ and capacitance $C$ encompassing its charging and inductive energy
\begin{equation}\label{eq:HLC}
    H_0 = \frac{Q^2}{2C}+\frac{\Phi^2}{2L} 
\end{equation}
where $Q$ is the charge accumulated in the capacitor and $\Phi$ is the magnetic flux traversing the inductor, serving as a pair of canonical conjugate variables, which 
are then promoted to noncommuting observables obeying the commutation relation
$[\hat{\Phi},\hat{Q}] = i$.

Introducing the standard annihilation and creation operators,
\begin{equation}
    \hat a=\frac{1}{\sqrt{2 Z_r}}(\hat \Phi-i Z_r\hat Q),\qquad
    \hat a^\dagger=\frac{1}{\sqrt{2Z_r}}(\hat \Phi+i Z_r\hat Q)
\end{equation}
with $Z_r=\sqrt{L/C}$ being the characteristic impedance, the previous Hamiltonian takes the usual form
\begin{equation}
    H_0 =\omega_r \left(\hat{a}^\dagger \hat{a}+\frac{1}{2}\right),
\end{equation}
where $\omega_r = 1/\sqrt{LC}$ is the resonant frequency.

Superconducting qubits, known as artificial atoms, are anharmonic oscillators in the form of non-linear LC circuits.
The nonlinearity is contributed by replacing the geometric inductance $L$ of the LC oscillator with a single Josephson junction, which is a non-dissipative and nonlinear inductor, or two parallel junctions forming a superconducting quantum interference device (SQUID).
The Hamiltonian of the circuit is modified by the presence of the Josephson junction taking the place of the linear inductor
\begin{equation}\label{eq:HJ}
    H_0=\frac{\hat Q^2}{2C}-E_J\cos\left(\frac{2\pi}{\Phi_0}\hat \Phi\right)=4E_C\hat n^2-E_J\cos\hat \varphi
\end{equation}
with $\Phi_0=h/2e$ the flux quantum and $E_J$ the Josephson energy, which is fixed for a single Josephson junction and flux-tunable for a SQUID. We also define the charge number operator $\hat n=\hat Q/2e$, the phase operator $\hat\varphi=(2\pi/\Phi_0)\hat\Phi$, and the charging energy $E_C=e^2/2C$.

In the transmon regime ($E_J/E_C\gg 1$)~\cite{koch2007}, the variance of the phase is relatively small $\Delta\hat\varphi\ll 1$, therefore the last term of Eq.~\eqref{eq:HJ} can be truncated to its first nonlinear correction
\begin{equation}\label{eq:HJ1}
    H_0\approx 4E_C\hat n^2+\frac12E_J\hat\varphi^2-\frac{1}{4!}E_J\hat\varphi^4
\end{equation}
with the first two terms corresponding to an LC circuit of
capacitance $C$ and inductance $E_J^{-1}(\Phi_0/2\pi)^2$, which can be expressed by the annihilation and creation operators
\begin{equation}
    \hat \varphi =\left(\frac{2E_C}{E_J}\right)^\frac14 (\hat a^\dagger+\hat a),\qquad 
    \hat n =\frac{i}{2}\left(\frac{E_J}{2E_C}\right)^\frac14 (\hat a^\dagger-\hat a)
\end{equation}

Making use of the rotating-wave approximation finally gives the Hamiltonian  
\begin{equation}\label{eq:HJ2}
    H_0\approx\omega_r \hat a^\dagger \hat a-\frac{E_C}{2}\hat a^\dagger \hat a^\dagger \hat a \hat a
\end{equation}
with $\omega_r=2\sqrt{2E_CE_J}-E_C$ the transition frequency between the ground and first excited states, and the last term of Eq. \eqref{eq:HJ2} denotes a Kerr nonlinearity.

Transmon qubits have a strong coupling to electric fields~\cite{Chen:2022quj}.
In the presence of an effective current density, the observed electric field generated inside an electromagnetic shield is determined with the standard methods for driven cavities as in Sec.~\ref{sec:rc}~\cite{Chaudhuri:2014dla}, which is given by
\begin{equation}\label{eq:Eeff}
    \bar E_{\rm eff}=\kappa \frac{\bar J_{\rm eff}}{\omega}
\end{equation}
with $\omega$ the driving frequency of the effective current, $\bar J_{\rm eff}$  the average current density within the shield, and an enhancement factor 
\begin{equation}
    \kappa=\left|\sqrt{V}\sum_n \frac{\omega^2}{\omega_n^2-\omega^2}G_n\vec \epsilon_n\cdot\hat n_C\right|
\end{equation}
where $n$ runs over all resonant modes, $\omega_n, G_n, \vec \epsilon_n$ are the the resonant frequency, geometric factor and mode function respectively, $\hat n_C$ is the normal vector of the conductor plate. 
On the other hand, we simply have $\kappa=1$ for the unshielded case.

The interaction Hamiltonian is given by
\begin{equation}\label{eq:HJ0}
    H_{\rm int}=\hat Q\bar E_{\rm eff} d=\frac{i}{\sqrt{2}\omega}\sqrt{C\omega_r}\kappa {\bar J_{\rm eff}}d\hat a^\dagger e^{-i\omega_r t}+h.c.,
\end{equation}
where $d$ is the effective distance between two conductor plates.

\subsection{Trapped ion}

Our analysis applies not only to electromagnetic resonant systems, but also generalizes to other kinds of quantum sensors, e.g., mechanical oscillators.
The motion of an ions in a Penning trap or Paul trap can be described in terms of three independent harmonic oscillator modes, realizing the mechanical oscillator. The free Hamiltonian given by
\begin{equation}
    H_0=\sum_n \omega^n_r \left(\hat a^\dagger_n\hat a_n+\frac{1}{2}\right)
\end{equation}
with $\omega^n_r$ the mechanical resonant frequency of the $n$-th vibrational mode. 
Mechanical oscillators such as trapped ions can serve as a high-Q resonator to measure small displacements due to weak forces and electromagnetic fields. The interaction Hamiltonian between the vibrational modes and the electric field induced by an effective current is 
\begin{equation}\label{eq:Hion}
    H_{\rm int}=e\vec E_{\rm eff} \cdot \hat{\vec{x}}=\sum_n\frac{e E_{\rm eff}^n}{\sqrt{2m_{\rm ion}\omega^n_r}}\hat a_n^\dagger e^{-i\omega^n_r t}+h.c..
\end{equation}
where $\hat{\vec{x}}$ is the three dimensional position operator, $\vec E_{\rm eff}$ is the induced electric field at the location of the ion, $m_{\rm ion}$ is the ion mass.

\section{Dark matter signals}\label{sec:dmh}

The effective current $J^\mu_{\rm eff}$ can be read off directly by rearranging the interaction Lagrangian between bosons and the electromagnetic field into the configuration $A_\mu J^\mu_{\rm eff}$.

The interaction between a quantum chromodynamic axion or an axion-like particle $a$ and a photon $A^\mu$ is described by the Lagrangian
\begin{equation}
    \mathcal{L}_a\supset \frac{1}{4} g_{a \gamma} a F_{\mu \nu} \tilde{F}^{\mu \nu},
\end{equation}
indicating the effective current 
\begin{equation}
    J_{\text{eff}}^{a\,\mu} = -g_{a\gamma} \tilde{F}^{\mu \nu} \partial_{\nu} a
\end{equation}
where $g_{a \gamma}$ is the axion-photon coupling, $F_{\mu \nu}$ and $\tilde{F}^{\mu \nu}$ denote the field-strength tensor and the dual tensor for the photon. In non-relativistic scenarios, the axion DM is uniform in space and oscillates at its Compton frequency corresponding to its mass $m_a$:
\begin{equation}
    a(\vec x,t)=a_0 \cos(m_a t+\phi_a)
\end{equation}
where $a_0$ is the oscillation amplitude and $\phi_a$ is a random phase.
With a background magnetic field $\vec B_0$, the effective current takes the form
\begin{equation}\label{eq:Ja}
    \vec{J}^a_{\rm eff} = g_{a \gamma} \partial_t a \vec B_0
\end{equation}

The kinetic mixing between dark photon $A'^\mu$ of mass $m_{A'}$ and the standard model photon $A^\mu$ is described by the Lagrangian in the interaction basis
\begin{equation}
    \mathcal{L}_{A'}\supset \epsilon m^2_{A^\prime} A^{\prime \mu} A_{\mu},
\end{equation}
denoting the effective current 
\begin{equation}\label{eq:JA}
    J_{\text{eff}}^{A^\prime\,\mu} = \epsilon m^2_{A^\prime} A^{\prime \mu},
\end{equation}
where $\epsilon$ is the the kinetic mixing coefficient. In the non-relativistic limit, the temporal component of dark photon is suppressed by $v_{A'}\sim 10^{-3}\ll 1$, while the spatial components are given by 
\begin{equation}
    \vec A'(\vec x,t)=A'_0\cos(m_{A'}t+\phi_{A'})\vec\epsilon_{A'}
\end{equation}
where $A'_0,\phi_{A'},\vec\epsilon_{A'}$ are the oscillation amplitude, phase and polarization vector of the dark photon.
One important difference from the axion case is that the effective current in Eq. \eqref{eq:JA} solely dependends on the dark photon field, thus an electromagnetic background field is not required. Specifically, the spatial component of the current is directly proportional to $\vec A'$.

From now on, we use a unified symbol 
\begin{equation}\label{eq: DMP}
    \Psi=\Psi_0 \cos(m_\Psi t+\phi_\Psi)
\end{equation}
to represent the bosonic DM field, where $\Psi=a$ or $A'$ represents the case of axion or dark photon with amplitude $\Psi_0$, Compton frequency $m_\Psi$ and a random phase $\phi_\Psi$.
As shown in Sec. \ref{sec:detect}, the spatial component of the effective currents due to DM couples to the resonance mode of a quantum sensor through the following Hamiltonian expression:
\begin{equation}\label{eq:hintA}
    {H}_{\alpha}= \alpha \Psi \left(\hat{a}\, e^{\mi \omega_r t}+\hat{a}^\dagger\, e^{-\mi \omega_r t}\right)/\sqrt{2}.
\end{equation}
where the coupling coefficients $\alpha$ are derived in accordance with Eqs. (\ref{eq:cavalp}, \ref{eq:HJ0}, and \ref{eq:Hion}), respectively:
\begin{equation}
    \alpha_{\textrm{cav}}\Psi=\sqrt{\frac{V}{\omega_r }} G \bar{J}_{\rm eff}, 
    \qquad   \alpha_{\textrm{JJ}}\Psi=\sqrt{C\omega_r}\bar E_{\rm eff}d
    \qquad   \alpha_{\textrm{ion}}\Psi=\frac{e \bar E_{\rm eff}}{\sqrt{m_{\rm ion}\omega^n_r}}
\end{equation}

The average current density $\bar{J}_{\rm eff}$, can be inferred from Eq. (\ref{eq:Ja}) for axion and Eq. (\ref{eq:JA}) for dark photon, respectively. The magnetic flux $\Phi_\Psi$ and electric field $\bar E_{\rm eff}$ are related to the effective current according to Eq. \eqref{eq:Phi} and Eq. \eqref{eq:Eeff}, respectively. The values of parameter $\alpha$ for each source and detection scheme are listed in Table.\,\ref{table:alpha}.

\begin{table*}[htb]
\begin{center}
\begin{tabular}{|c||*{2}{c|}}\hline
\diaghead{\theadfont HaloscopeSource}
{Haloscope}{Source} & Axion & Dark Photon \\\hline\hline
Cavity & $g_{a \gamma}G B_0 m_a\sqrt{{V}/{\omega_r}} $
& $\epsilon G m^2_{A^\prime}\sqrt{{V}/{\omega_r}}$ \\\hline
Transmon & $g_{a \gamma} \kappa d B_0\sqrt{C\omega_r}$
& $\epsilon\kappa d m_{A^\prime}\sqrt{C\omega_r}$ \\\hline
Trapped ion & $g_{a \gamma}\kappa  B_0 e/\sqrt{m_{\rm ion}\omega^n_r}$
& $\epsilon\kappa e m_{A^\prime}/\sqrt{m_{\rm ion}\omega^n_r}$ \\\hline
\end{tabular}
\caption{Effective couplings $\alpha$ between bosons and sensor modes in Eq.\,(\ref{eq:hintA}). \label{table:alpha}}
\end{center}
\end{table*}

According to Eqs. (\ref{eq: DMP}-\ref{eq:hintA}), the DM field couples resonantly to the sensor when $\omega_r=m_\Psi$. Making use of the rotating wave approximation, a harmonic quantum sensor initialized in its ground state evolves into a coherent state
\begin{equation}\label{eq:har}
    |\beta\rangle = \hat{\mathcal{D}}(\beta)|0\rangle=e^{-\frac{1}{2}|\beta|^2}\sum_n \frac{\beta^n}{\sqrt{n!}}|n\rangle,
\end{equation}
where $\hat{\mathcal{D}}(\beta)=\exp({\beta\hat a^\dagger-\beta^*\hat a})$ is a displacement operator, $\beta=-\frac{i}{2\sqrt{2}}\alpha\Psi_0e^{i\phi_\Psi}\tau$ is the amplitude of the displacement, $\tau<\min\{T_\Psi, T_s\} $ is the signal integration time which is required to be within the coherence time of the DM $T_\Psi$ and the sensor $T_s$. 
While for an anharmonic sensor, the Hamiltonian leads to a Rabi oscillation between the ground and first excited states. Thus the ground state evolves into 
\begin{equation}\label{eq:anhar}
    |\beta\rangle=\cos|\beta||g\rangle+e^{i\phi_\Psi}\sin|\beta||e\rangle.
\end{equation}

Since the DM signal is extremely weak, the resulting average occupation number of the sensor mode is much less than one ($|\beta|^2\ll 1$). Thus we can safely restrict our attention to the first two energy levels and expand the state of sensor \eqref{eq:har} and \eqref{eq:anhar} according to $\beta$, which give the same result as follows
\begin{equation}
    |\beta\rangle\simeq \left(1-\frac{|\beta|^2}{2}\right)|g\rangle+\beta|e\rangle+O(\beta^3)
\end{equation}

The DM signal in an ideal quantum sensor is signified by the expectation value of the observable $\mathcal{O}\equiv|e\rangle\langle e|$, i.e., the excitation probability:
\begin{equation}
P_e=\langle \beta|\mathcal{O}|\beta\rangle\simeq|\beta|^2.
\end{equation}

\section{Qubit noise models}\label{sec:qnm}

In the main text we mainly addressed three kinds of error sources: (i) hardware infidelities in the form of depolarizing Pauli error, (ii) state preparation and measurement (SPAM) errors, and (iii)  decoherence in the form of thermal relaxation/excitation and dephasing errors. In this section we discuss the quantum channels that we use to model each error.

The noise acting on a qubit can be described by a trace-preserving quantum operation $\mathcal{E}$, which can be expanded in an operator-sum representation with operation elements $\{E_i\}$. The density matrix of the qubit $\rho$ evolves under the acting of noise as
\begin{equation}
    \rho\rightarrow \mathcal{E}(\rho)=\sum_i E_i \rho E_i^\dagger
\end{equation}

The combination of two error channels $\mathcal{E}_{AB}=\mathcal{E}_A\circ\mathcal{E}_B$ where $\mathcal{E}_A(\rho)=\sum_i A_i \rho A_i^\dagger$ and $\mathcal{E}_B(\rho)=\sum_j B_j \rho B_j^\dagger$
is given by
\begin{equation}\label{combine}
    \mathcal{E}_{AB}(\rho)=\mathcal{E}_A(\mathcal{E}_B(\rho))=\sum_{ij} (A_iB_j) \rho (A_iB_j)^\dagger
\end{equation}

The Choi-matrix~\cite{Blank:2020zfq} provides an alternative representation for a quantum channel. For a single qubit, the Choi-matrix of the quantum channel $\mathcal{E}$ is defined as
\begin{equation}
    C^{(1)}=\sum_{i,j}|i\rangle\langle j|\otimes\mathcal{E}(|i\rangle\langle j|)
\end{equation}

The evolution of the density matrix $\rho^{(1)}$ with respect to the Choi matrix $C^{(1)}$ can be described as
\begin{equation}\label{che}
    \rho^{(1)}\rightarrow\mathcal{E}(\rho^{(1)})=\mathrm{Tr}_1[C^{(1)}(\rho^{(1)T}\otimes I)]
\end{equation}
where $T$ represents transpose and $\mathrm{Tr}_1$ is the partial trace over subsystem 1 in which the density matrix $\rho^{(1)}$ resides.

For multipule qubits the Choi matrix $C^{(n)}$ can be constructed similarly by the tensor products.

\subsection{The depolarizing channel}

The depolarizing channel simulates the bit-flip and phase-flip errors due to operation infidelities within a certain process, which are represented by the following Pauli operators
\begin{equation}
    D_0=\sqrt{1-\frac{3}{4}\lambda}I_2,\quad D_1=\frac{\sqrt{\lambda}}{2}\sigma_x,\quad
    D_2=\frac{\sqrt{\lambda}}{2}\sigma_y,\quad D_3=\frac{\sqrt{\lambda}}{2}\sigma_z
\end{equation}

The noise operation on a single qubit $\rho^{(1)}$ gives
\begin{equation}
    \mathcal{D}(\rho^{(1)})=\sum_{i=0}^3 D_i \rho D_i^\dagger= (1-\lambda)\rho^{(1)}+\frac{\lambda}{2}I
\end{equation}

For the example we considered in the main text, the depolarizing channel $\mathcal{D}_{AB}$ was devised to apply on sensor $A$ and $B$ with different parameters $\lambda_A$ and $\lambda_B$, where the two-qubit system evolves as
\begin{equation}
    \rho'^{(2)}\rightarrow\mathcal{D}_{AB}(\rho'^{(2)})=\sum_{i=0}^3 \sum_{j=0}^3 (D_i^A D_j^B) \rho'^{(2)} (D_i^A\otimes D_j^B)^\dagger
\end{equation}
The expectation value becomes
\begin{equation}
    Tr[\mathcal{D}_B(\rho'^{(2)})\cdot\mathcal{O}'^{(2)}_{P}]=\frac{\lambda_B-\lambda_A}{2}(1-p_{reset}^Ap_{1}^A-p_{reset}^Bp_{1}^B)+2p_{T_2}^Ap_{T_2}^B|\beta|^2\left(1-\frac{\lambda_A+\lambda_B}{2}+\frac{p_{T_1}^A+p_{T_1}^B}{4p_{T_2}^Ap_{T_2}^B}(\lambda_A-\lambda_B)\right),
\end{equation}
under the influence of $\mathcal{D}_{AB}$, with background noise of order $O(\lambda)$ added.

In the contrast, the symmetric depolarizing channel applying to $n$-qubits with parameter $\lambda$ gives 
\begin{equation}
    \mathcal{D}(\rho^{(n)})=(1-\lambda)\rho^{(n)}+\frac{\lambda}{2^n}I
\end{equation}
\begin{equation}
    Tr[\mathcal{D}(\rho'^{(N)})\cdot\mathcal{O}'^{(N)}_{P}]=(1-\lambda)Tr[\rho'^{(N)}\cdot\mathcal{O}'^{(N)}_{P}],
\end{equation}
which only leads to a $(1-\lambda)$ suppression on the signal strength without adding additional noises.

\subsection{State Preparation and Measurement (SPAM) Channel}

The SPAM quantum channel can be simply represented by a simple Pauli X error with the following Kraus operators
\begin{equation}
    S_0=\sqrt{1-p_2}I,\quad S_1=\sqrt{p_2}X
\end{equation}
where $p_2$ is the is the probability that the measurement (state preparation) is incorrect.
The effect of the SPAM channel can be expressed as
\begin{equation}
    \mathcal{S}(\rho^{(1)})=S_0\rho^{(1)}S_0+S_1\rho^{(1)}S_1
\end{equation}

\subsection{Thermal Decoherence and Dephasing Channel}
The single qubit thermal relaxation channel is parameterized by relaxation time constants $T_1$ and $T_2$ with relation $T_2\leq 2T_1$, gate time $t$, and excited state thermal population $p_1$. 

If $T_2\leq T_1$, the error can be expressed as a mixed reset and Pauli $Z$ error channel, where the single qubit reset quantum error channel is given by the map
\begin{equation}
    \rho^{(1)}\rightarrow \mathcal{R}(\rho^{(1)})=(1-p_1)|0\rangle\langle 0|+p_1|1\rangle\langle 1|.
\end{equation}
The effect of the thermal decoherence and dephasing channel can be expressed as
\begin{equation}
    \rho^{(1)}\rightarrow \mathcal{E}(\rho^{(1)})=\frac{p_{T_1}+p_{T_2}}{2}\rho^{(1)}+\frac{p_{T_1}-p_{T_2}}{2}Z\rho^{(1)}Z+p_{\mathrm{reset}}\mathcal{R}(\rho^{(1)})
\end{equation}
where $p_{T_1}=e^{-t/T_1}$ and $p_{T_2}=e^{-t/T_2}$ are the probability for a qubit to relax and dephase during time $t$, $p_{\mathrm{reset}}=1-p_{T_1}$ is the probability for a qubit to reset to an equilibrium state.

While for $T_1\leq T_2 \leq 2T_1$, the error can only be expressed by the following Choi matrix
\begin{equation}
    C^{(1)}=\sum_{i,j}|i\rangle\langle j|\otimes\mathcal{E}(|i\rangle\langle j|)=\begin{pmatrix}
        1-p_1p_{\mathrm{reset}}&0&0&p_{T_2}\\
        0&p_1p_{\mathrm{reset}}&0&0\\
        0&0&(1-p_1)p_{\mathrm{reset}}&0\\
        p_{T_2}&0&0&1-(1-p_1)p_{\mathrm{reset}}
    \end{pmatrix}
\end{equation}
where
\begin{align}
    \mathcal{E}(|0\rangle\langle 0|)&=(1-p_1p_{\mathrm{reset}})|0\rangle\langle 0|+p_1p_{\mathrm{reset}}|1\rangle\langle 1|\\
    \mathcal{E}(|0\rangle\langle 1|)&=p_{T_2}|0\rangle\langle 1|\\
    \mathcal{E}(|1\rangle\langle 0|)&=p_{T_2}|1\rangle\langle 0|\\
    \mathcal{E}(|1\rangle\langle 1|)&=(1-p_1)p_{\mathrm{reset}}|0\rangle\langle 0|+(1-(1-p_1)p_{\mathrm{reset}})|1\rangle\langle 1|
\end{align}
And the evolution of the density matrix $\rho^{(1)}$ with respect to
the Choi matrix $C^{(1)}$ is given by Eq. \eqref{che}. One can show that the two representations are equivalent to each other when $T_2\leq T_1$.

For multipule qubits, the error channel can be constructed following Eq. \eqref{combine}.  
The Choi matrix is constructed as
\begin{equation}
    C^{(2)}=\sum_{i_Ai_Bj_Aj_B}|i_Ai_B\rangle\langle j_Aj_B|\otimes\mathcal{E}_{AB}(|i_Ai_B\rangle\langle j_Aj_B|)=\sum_{i_Ai_Bj_Aj_B}|i_A\rangle\langle j_A|\otimes|i_B\rangle\langle j_B|\otimes\mathcal{E}_A(|i_A\rangle\langle j_A|)\otimes(\mathcal{E}_B|i_B\rangle\langle j_B|)
\end{equation}
then 
\begin{equation}
    \rho^{(2)}\rightarrow\mathcal{E}_{AB}(\rho^{(2)})=\mathrm{Tr}_1[C^{(2)}(\rho^{(2)T}\otimes I_4)]
\end{equation}

The expectation value under the influence of thermal decoherence and dephasing  error becomes
\begin{equation}
Tr[\mathcal{E}_{AB}(\rho'^{(2)})\cdot\mathcal{O}'^{(2)}_{P}]
=\frac{\epsilon_A + \epsilon_B}{2}(p_{1}^B-p_{1}^A)+\frac{\epsilon_B-\epsilon_A}{2}(p_{T_1}^Ap_{1}^A+p_{T_1}^Bp_{1}^B)
+\left(1-\frac{\epsilon_A + \epsilon_B}{2}-\frac{\epsilon_A-\epsilon_B}{4}\frac{p_{T_1}^A+p_{T_1}^B}{p_{T_2}^Ap_{T_2}^B}\right)2p_{T_2}^Ap_{T_2}^B|\beta|^2,\label{eq:ndeapp} 
\end{equation}
The background noise added due to the thermal relaxation channel is on an order of $O(\bar n_r\tau_g/T_1)$.

\section{Multiple sensors containing all coherence}\label{sec:diag}

For $N$ quantum sensors there are totally ${N(N-1)}/{2}$ coherence channels. 
In contrast of the particular observable we considered in the main text, the one containing all coherence channels can be defined as 
\begin{equation}\label{obsa}
    \mathcal{O}^{(N)}_{CA}=\sum_{I>J} \mathcal{O}^{(2)}_{C_{IJ}}=\sum_{I\neq J} |E^{(N)}_I\rangle\langle E^{(N)}_J|,
\end{equation}
which gives noise-free signals enhanced by a factor of $N(N-1)$:
\begin{equation}
    Tr[\rho^{(N)}\cdot\mathcal{O}^{(N)}_{CA}]=2\sum_{I>J}p_{T_2}^Ip_{T_2}^J|\beta|^2.
\end{equation}

It can be diagonalized in the same way by the unitary transformation $\mathcal{U}^{(N)}_C$:
\begin{equation}
\mathcal{O}'^{(N)}_{PA}=\mathcal{U}^{(N)}_C\mathcal{O}^{(N)}_{CA}\mathcal{U}^{(N)\dagger}_C=N|E^{(N)}_{N}\rangle\langle E^{(N)}_{N}|-\sum_{I=1}^{N} |E^{(N)}_I\rangle\langle E^{(N)}_I|.
\end{equation}

Although the observable contains more coherence channels than $\mathcal{O}^{(N)}_{C}$, the background noise for the measurement of $\mathcal{O}^{(N)}_{CA}$ would increase as $O(N^2)$ instead of $O(N)$ if we consider the hardware infidelities in implementations of $\mathcal{U}^{(N)}_C$.
This is because errors due to all steps of $\mathcal{U}^{(N)}_{C_{IJ}}$ will contribute to the final measurement result of $\mathcal{O}^{(N)}_{CA}$, while for $\mathcal{O}^{(N)}_{C}$ only the last one does.

\begin{figure}[]
\centering
\includegraphics[width=0.3\textwidth]{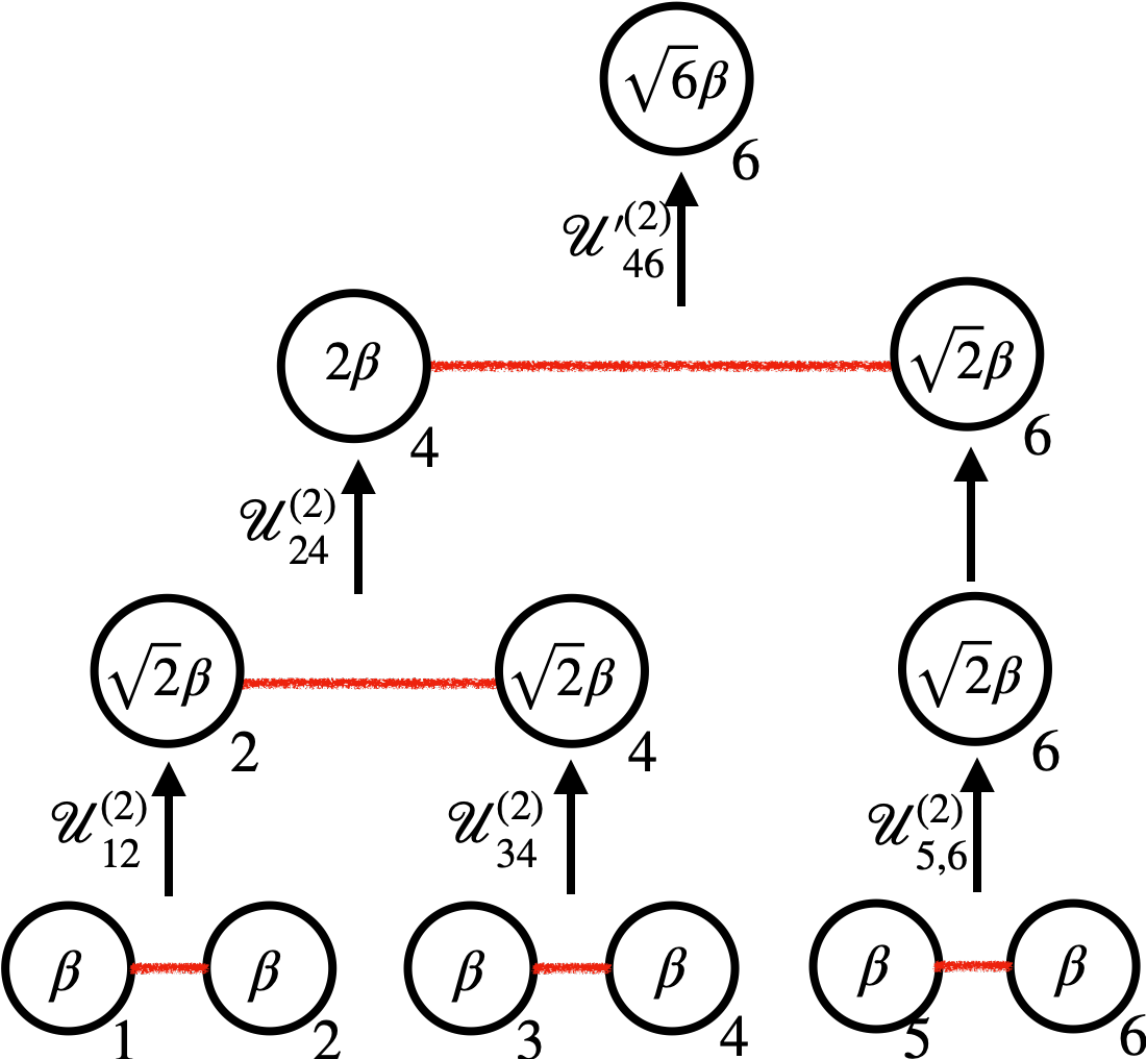}
\caption{Decomposition of $\mathcal{U}^{(6)}_C$.
}
\label{fig:img6}
\end{figure}
Note that $N$ does not necessarily be an order of 2. For any integer $N$, $\mathcal{U}^{(N)}_C$ can be decomposed into $N-1$ two-body unitary transformations and implemented within $\lceil\log_2 N\rceil$ steps with a complete binary tree structure. As an example for $N=6$, the unitary matrix $\mathcal{U}^{(6)}_C$ diagonalizing $\mathcal{O}^{(6)}_{C}$ or $\mathcal{O}^{(6)}_{CA}$ can be decomposed as following
\begin{equation}
\mathcal{U}^{(6)}_C=\mathcal{U'}^{(2)}_{C_{46}}\mathcal{U}^{(2)}_{C_{24}}\mathcal{U}^{(2)}_{C_{12}}\mathcal{U}^{(2)}_{C_{34}}\mathcal{U}^{(2)}_{C_{56}}
\end{equation}
as Fig. \ref{fig:img6} shows, where $\mathcal{U}^{(2)}_{C_{IJ}}$ is the same as defined in the main text, and
\begin{equation}
    \mathcal{U'}^{(2)}_{C_{46}}=\begin{pmatrix}
        1&0&0&0\\ 0&\frac{1}{\sqrt{3}}&\sqrt{\frac{2}{3}}&0\\ 0&-\sqrt{\frac{2}{3}}&\frac{1}{\sqrt{3}}&0\\ 0&0&0&1
    \end{pmatrix}.
\end{equation}

In general, we have
\begin{equation}
    \mathcal{U'}^{(2)}_{C_{IJ}}=\begin{pmatrix}
        1&0&0&0\\ 0&\cos\theta_{IJ}&\sin\theta_{IJ}&0\\ 0&-\sin\theta_{IJ}&\cos\theta_{IJ}&0\\ 0&0&0&1
    \end{pmatrix}
\end{equation}
where $\cos\theta_{IJ}=\frac{\beta_J}{\sqrt{\beta_I^2+\beta_J^2}}, \sin\theta_{IJ}=\frac{\beta_I}{\sqrt{\beta_I^2+\beta_J^2}}$, and $\beta_{I/J}$ is the amplitude of the excited state in sensor $I/J$.

\end{document}